\pdfoutput=1

\documentclass[12pt,a4paper,hyperref]{JHEP3}
\let\ifpdf\relax
\usepackage{ifpdf,color}
\let\normalcolor\relax
\usepackage{mathtools,cite}
\newcommand{\eq}[1]{\vspace{-0.5pt}\begin{equation}#1\vspace{-0.5pt}\end{equation}}

\newcommand{\fwbox}[2]{\text{\makebox[#1][c]{$\hspace{-150pt}\displaystyle#2\hspace{-150pt}$}}}
\newcommand{\fwboxL}[2]{\text{\makebox[#1][l]{$#2$}}}
\newcommand{\fwboxR}[2]{\text{\makebox[#1][r]{$#2$}}}
\newcommand{\fig}[3]{\raisebox{#1}{\ \includegraphics[scale=#2]{#3}}}
\newcommand{\mi}{\raisebox{0.75pt}{\scalebox{0.75}{$\,-\,$}}}
\newcommand{\pl}{\raisebox{0.75pt}{\scalebox{0.75}{$\,+\,$}}}
\renewcommand{\phi}{\varphi}
\newcommand{\m}[1]{(#1)}
\newcommand{\newcap}{\mathrm{\raisebox{1.75pt}{\scalebox{0.6}{{$\,\bigcap\,$}}}}}
\newcommand{\inter}[2]{(\hspace{-0.5pt}#1\hspace{-0.5pt})\hspace{-0.5pt}\newcap\hspace{-0.5pt}(\hspace{-0.5pt}#2\hspace{-0.5pt})}
\newcommand{\interBig}[2]{(\hspace{-0.5pt}#1\hspace{-0.5pt})\hspace{-0.5pt}\raisebox{-1pt}{\scalebox{1.4}{$\newcap$}}\hspace{-0.5pt}(\hspace{-0.5pt}#2\hspace{-0.5pt})}
\newcommand{\asb}[4]{\langle\hspace{-1pt}#1\hspace{-0.5pt}|\hspace{-0.5pt}#2\hspace{-0.5pt}\pl\hspace{-0.5pt}#3\hspace{-0.5pt}|\hspace{-0.5pt}#4]}
\newcommand{\ab}[1]{\langle\hspace{-0.5pt}#1\hspace{-0.5pt}\rangle}
\renewcommand{\sb}[1]{[\hspace{-0.5pt}#1\hspace{-0.5pt}]}
\renewcommand{\hat}{\widehat}

\newcommand{\bigger}[1]{\raisebox{-2.25pt}{\scalebox{1.75}{$#1$}}}
\newcommand{\Pl}{Pl\"ucker }

\definecolor{hblue}{rgb}{0,0,0.75}
\definecolor{hred}{rgb}{0.65,0,0.15}
\definecolor{dim}{rgb}{0.6,0.6,0.6}

\preprint{CCNY-HEP-16-06}
\title{{\LARGE \mbox{Stratifying On-Shell Cluster Varieties: the}}\\ {\LARGE\mbox{Geometry of Non-Planar On-Shell Diagrams}}}
\author{{\normalsize \mbox{Jacob~L.~Bourjaily$^1$, Sebasti\'an~Franco$^{2,3}$, Daniele~Galloni$^4$ and Congkao~Wen\mbox{$^{5}$}}}\\
\mbox{{\mbox{$^1$}\ Niels Bohr International Academy and Discovery Center, University of Copenhagen,}}\\
\mbox{{\mbox{$^{\phantom{1}}$}\ Blegdamsvej 17, DK-2100 Copenhagen \O, Denmark}}\\
\mbox{{\mbox{$^2$}\ Department of Physics, The City College of the CUNY}}\\
\mbox{{\mbox{$^{\phantom{2}}$}\ 160 Convent Avenue, New York, NY 10031, USA}}\\
\mbox{{\mbox{$^3$}\ The Graduate School and University Center, The City University of New York}}\\
\mbox{{\mbox{$^{\phantom{3}}$}\ 365 Fifth Avenue, New York NY 10016, USA}}\\
\mbox{{\mbox{$^4$}\ I.N.F.N.\ Sezione di Torino, Via Pietro Giuria 1, 10125 Torino, Italy}}\\
\mbox{{\mbox{$^5$}\ I.N.F.N.\ Sezione di Roma ``Tor Vergata"}}\\
\mbox{{\mbox{$^{\phantom{5}}$}\ Via della Ricerca Scientifica, 00133 Roma, Italy}}
}

\date{\today}
\abstract{The correspondence between on-shell diagrams in maximally supersymmetric Yang-Mills theory and cluster varieties in the Grassmannian remains largely unexplored beyond the planar limit. In this article, we describe a systematic program to survey such `on-shell varieties', and use this to provide a complete classification in the case of $G(3,6)$. In particular, we find exactly 24 top-dimensional varieties and 10 co-dimension one varieties in $G(3,6)$---up to parity and relabeling of the external legs. We use this case to illustrate some of the novelties found for non-planar varieties relative to the case of positroids, and describe some of the features that we expect to hold more generally.
}

\begin{document}
\newpage
\section{Introduction}\label{sec:introduction}\vspace{-6pt}
In recent years there has been tremendous progress in our understanding of scattering amplitudes in quantum field theory. The developments have been most notable for maximally supersymmetric ($\mathcal{N}\!=\!4$) Yang-Mills theory (SYM) (see e.g.\ \cite{ArkaniHamed:book,Elvang:2013cua,Henn:2014yza} and references therein). New discoveries include results at high loop level \cite{Goncharov:2010jf,Golden:2013xva,Golden:2014xqa,Golden:2014pua,Drummond:2014ffa,Dixon:2011pw,Dixon:2011nj,Dixon:2014voa,Dixon:2014iba, Basso:2013vsa,Basso:2013aha,Basso:2014koa,Huang:2013owa,Huang:2014xza,Bourjaily:2015bpz}, the existence of infinite dimensional Yangian symmetries \cite{Drummond:2009fd} as a result of combining superconformal and dual superconformal invariance \cite{Alday:2007hr,Drummond:2006rz}, and integrability \cite{Beisert:2003yb,Beisert:2006ez}. 

These advances have motivated a new formulation of the foundations of field theory that incorporates these structures at a fundamental level. The new approach relies on novel mathematical and geometrical objects such as the Grassmannian \cite{P,ArkaniHamed:2009dn,ArkaniHamed:2009vw,Kaplan:2009mh,Mason:2009qx,ArkaniHamed:2009dg}, on-shell diagrams \cite{ArkaniHamed:2012nw} and the amplituhedron \cite{Arkani-Hamed:2013jha,Arkani-Hamed:2013kca,Franco:2014csa,Lam:2014jda,Bai:2014cna,Arkani-Hamed:2014dca,Galloni:2016iuj}. So far, most of the new results have been confined to the planar limit of $\mathcal{N}\!=\!4$ SYM. It is natural to ask whether, and if so how, these ideas extend beyond the planar sector. With this goal in mind, the study of non-planar on-shell diagrams \cite{Arkani-Hamed:2014bca,Franco:2015rma,Chen:2015bnt,Chen:2015lyz,Chen:2015qna} and the non-planar amplituhedron \cite{Bern:2015ple} has been recently initiated.

The all-loop integrand in planar $\mathcal{N}\!=\!4$ SYM can be constructed {\it directly} in terms of on-shell diagrams \cite{ArkaniHamed:2010kv,ArkaniHamed:2012nw}. While at present it is not known whether non-planar on-shell diagrams provide sufficient building blocks for directly expressing non-planar scattering amplitudes in the same way, they clearly encode {complete} information about the theory---sufficient to reconstruct amplitudes at all orders using (generalized) unitarity \cite{Bern:1994cg,Bern:1994zx} (see also \cite{ArkaniHamed:2010gh,Bourjaily:2013mma,Bourjaily:2015jna} for more recent developments). At least for $\mathcal{N}\!=\!4$ SYM, it has long been known that only a {\it finite number of distinct on-shell functions exist}: all-order information is captured by a finite (and small) number of elementary objects. And yet (beyond the case of MHV amplitudes \cite{Arkani-Hamed:2014bca}), very little is known about their scope or relations beyond the planar limit.

We seek to improve our understanding of these functions through a systematic survey of concrete examples. We describe a general procedure to explore the space of functions for general amplitudes, and we use this to completely classify the on-shell functions relevant to the 6-particle NMHV amplitude. This program is made possible through the correspondence between on-shell functions and cluster sub-varieties of Grassmannian manifolds described in \mbox{ref.\ \cite{ArkaniHamed:2012nw}}. We refer to the cluster varieties associated with on-shell functions as {\it on-shell varieties}.

There remain many important and fundamental open questions about non-planar on-shell varieties. In this work, we begin to answer some of these questions through a complete classification in the case of $G(3,6)$. We will find that much of the simplicity of positroids (the on-shell varieties of planar on-shell functions) is preserved, but that important new features also arise. We illustrate these novelties with examples from $G(3,6)$, and describe what aspects we expect to be preserved more generally.

This work is organized as follows. In \mbox{section \ref{sec:review_of_onshell_diagrams}} we provide a lightning review of the basic correspondence between on-shell functions and Grassmannian geometry. We describe the map between on-shell diagrams and on-shell varieties and the basic operations from which each can be built. There exist two known equivalence transformations among on-shell diagrams that leave their on-shell functions and the corresponding on-shell varieties unchanged; these are cluster mutations for the variety. All diagrams related to a planar (positroid) on-shell variety are related by these transformations alone. Is this true for non-planar varieties? We find that this continues to be true for at least $G(3,6)$.  

In \mbox{section \ref{sec:stratification_of_top_forms}}, we describe how the space of on-shell varieties can be surveyed by direct construction of representative on-shell diagrams. We describe two systematic approaches to building non-planar on-shell diagrams from planar ones: by attaching {\it BCFW bridges}; and by gluing additional external legs. Of these, only the latter is truly general. While representatives of all planar on-shell diagrams can be obtained through a sequence of BCFW bridges, we find that this fails in general. In particular, we find exactly two 9-dimensional on-shell varieties of $G(3,6)$ that cannot be obtained through a sequence of bridges. These varieties are represented by the diagrams,
\eq{\fwboxR{0pt}{\fig{-47.25pt}{0.8}{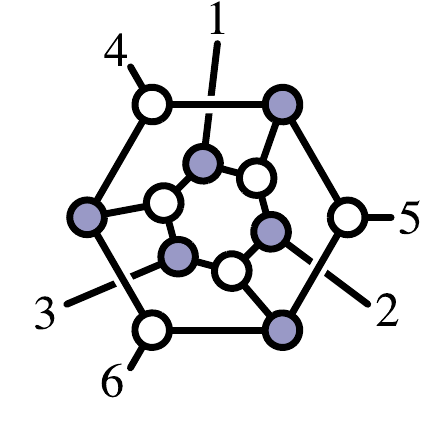}\hspace{10pt}}\quad\mathrm{and}\quad\fwboxL{0pt}{\fig{-41.1pt}{0.8}{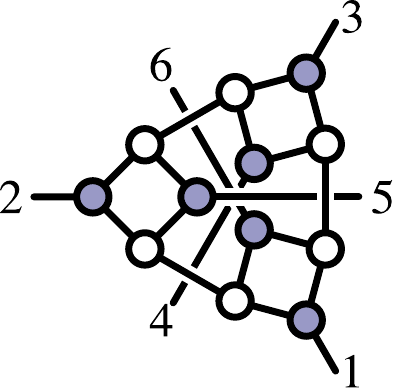}}\label{non_bridge_constructible_graphs}}
In addition to these, we find 22 other top-dimensional varieties (all bridge constructible). These examples, and some of their novel features are described in more detail in \mbox{section \ref{sec:g36_classification}}. In \mbox{Appendix \ref{appendix:top_forms}}, we give representative diagrams for all 24 top-dimensional varieties that exist, together with their volume forms expressed in Pl\"{u}cker coordinates. We also provide representatives of all (10) co-dimension one on-shell varieties (corresponding to {\it leading singularities}) in \mbox{Appendix \ref{appendix:leading_singularities}}. For each representative leading singularity, we have given an explicit formula for the corresponding on-shell function in terms of spinor-helicity variables. In \mbox{Appendix \ref{appendix:higher_singularities}}, we outline the classification of the higher co-dimension on-shell varieties of $G(3,6)$---all of which are positroid varieties below dimension $6$.

\newpage
\section{Grassmannian Representations of On-Shell Diagrams}\label{sec:review_of_onshell_diagrams}\vspace{-6pt}
There exists a fundamental and deep correspondence between {on-shell diagrams} in quantum field theory and cluster varieties in the Grassmannian. In this section, we sketch the essential ingredients of this story, illustrating all of the essential ideas required for the study of non-planar on-shell varieties for maximally supersymmetric ($\mathcal{N}\!=\!4$) Yang-Mills theory (SYM). The on-shell functions of SYM correspond to (generally non-planar) undirected graphs called {\it on-shell diagrams} constructed from two fundamental vertices,
\vspace{-26pt}\eq{\fig{-42.15pt}{1}{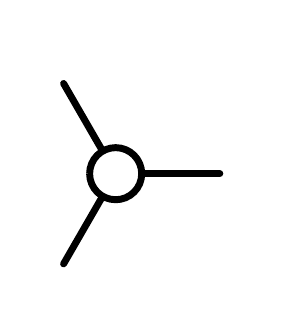}\text{and}\fig{-42.15pt}{1}{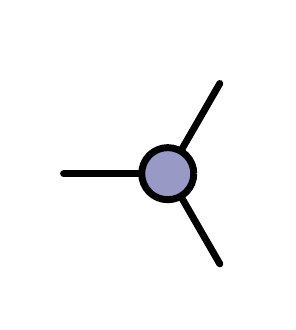}\label{three_point_vertices}\vspace{-6pt}}
~\\[-20pt]
\noindent which represent the three-particle amplitudes of $\mathcal{N}\!=\!4$ SYM. Examples of on-shell diagrams constructed from these vertices can be seen in (\ref{non_bridge_constructible_graphs}). Let $n$ denote the number of {\it external} legs of the diagram. For a trivalent diagram with $n_I$ internal edges, $n_B$ black vertices and $n_W$ white vertices, it is convenient to define
\vspace{-2pt}\eq{k\equiv 2n_B+n_W-n_I\,.\label{k_for_trivalent}\vspace{-2pt}}

We will refer to an on-shell diagram as {\it planar} if it admits an embedding on the disk without crossings and for which all external lines are along the boundary of the disk; a {\it non-planar} diagram is one that does not admit such an embedding. 

As described in \mbox{ref.\ \cite{ArkaniHamed:2012nw}}, for any on-shell diagram there exists a corresponding submanifold $C\!\subset\!G(k,n)$ of the Grassmannian of $k$-planes in $n$ dimensions. We will often consider this submanifold as being represented by a $k\times n$ matrix \mbox{$C\!\equiv\!(c_1,\ldots,c_n)$}, with the columns $c_a\!\in\!\mathbb{C}^k$ indexed by the labels of the external legs, $a\!\in\{1,\ldots,n\}$. (When the diagram is planar, it is natural to order the columns according to the diagram's plane embedding; but no preferred ordering exists for non-planar diagrams.) The on-shell diagram endows the submanifold $C$ with canonical (e.g.\ cluster) coordinates $\{\alpha_i\}$ and a volume-form $\Omega$. We refer to coordinates as being {\it canonical} if, when expressed in terms of them, $\Omega$ takes the form:
\vspace{-4pt}\eq{\Omega=\frac{d\alpha_1}{\alpha_1}\wedge\cdots\wedge\frac{d\alpha_d}{\alpha_d}\,.\label{volume_form_in_canonical_coordiantes}\vspace{-2.5pt}}
Notice that we have used `$d$' to denote the {\it number of cluster coordinates}. This number need not be equal to the dimensionality of the corresponding variety; there can be degeneracies among the coordinates. Nevertheless, the number of cluster coordinates is a useful characteristic for any graph $\Gamma$; for a graph with $n_V$ vertices,
\vspace{-2pt}\eq{d(\Gamma)=\underset{\text{(trivalent)}}{2n_V-n_I}=\underset{\text{(general valency)}}{n+n_I-n_V}=n_F-\chi,
\label{formula_for_d_of_a_graph}\vspace{-2pt}}
where $n_F$ counts the number of cycles (including boundaries along which external edges terminate) of an embedding of $\Gamma$ on a surface with Euler characteristic $\chi$.

\vspace{-6pt}\subsection{On-Shell Functions and Equivalence Relations}\label{subsec:on_shell_functions_and_moves}\vspace{-6pt}
The physical significance of this correspondence between Grassmannian geometry and on-shell diagrams follows from the fact that on-shell diagrams represent physically important functions called {on-shell functions}. The on-shell function $f_\Gamma$ associated with an on-shell diagram $\Gamma$ can be represented according to:
\vspace{-3pt}\eq{f_\Gamma\equiv\int\!\Omega_{C}\,\,\delta^{k\times4}\hspace{-0.5pt}\big(C(\alpha)\!\cdot\!\widetilde{\eta}\big)\delta^{k\times2}\hspace{-0.5pt}\big(C(\alpha)\!\cdot\!\widetilde\lambda\big)\delta^{2\times(n-k)}\hspace{-0.5pt}\big(\lambda\!\cdot\!C^{\perp}(\alpha)\big).\label{grassmannian_representation_of_on_shell_functions}\vspace{-3pt}}
The details of this formula need not concern us here. (The interested reader should refer to \mbox{ref.\ \cite{ArkaniHamed:book}}.) But this correspondence makes manifest two equivalence transformations among on-shell diagrams that leave the corresponding on-shell functions unchanged. The first of these transformations is fairly trivial: any chain of same-colored vertices can be arbitrarily rearranged---e.g.,\
\vspace{-3pt}\eq{\fwboxR{0pt}{\fig{-47.65pt}{1}{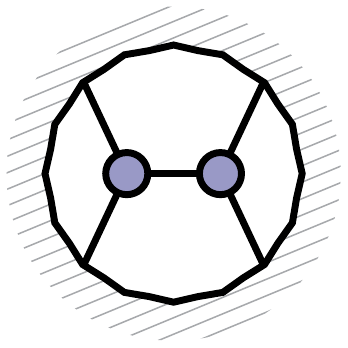}}\bigger{\Leftrightarrow}\fwbox{100pt}{\hspace{-3.5pt}\fig{-47.65pt}{1}{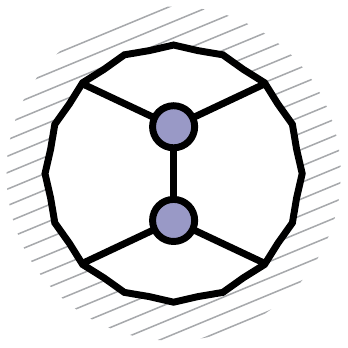}}\;\bigger{\Leftrightarrow}\fwboxL{0pt}{\hspace{-3.5pt}\fig{-47.65pt}{1}{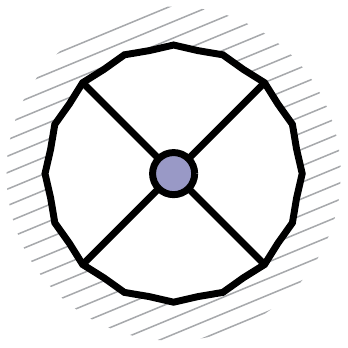}}\label{merge_expand_identity}\vspace{-3pt}}
This equivalence relation naturally suggests that we define higher-valency vertices so that all on-shell diagrams are made bipartite, trivializing the relation (\ref{merge_expand_identity}). (Bivalent vertices of either color can also be added to any edge without affect; this provides another way to render any on-shell diagram bipartite.) This is arguably the right thing to do, and greatly simplifies much of the analysis. Note, however, that several of the characteristics of diagrams quoted herein such as the formula for $k$ in (\ref{k_for_trivalent}) must be altered accordingly: for a bipartite graph involving $n_W$ white vertices and $n_{B}^v$ of black vertices with valency $v$, $k$ would given by:
\vspace{-3pt}\eq{k\equiv\Big(\sum_{v}(v\mi1)n_B^v\Big)+n_W-n_I\,.\label{k_for_general_valency}\vspace{-7pt}}

The second equivalence relation among on-shell diagrams is much more interesting. It is the so-called {\it square-move} \cite{Hodges:2005aj}:
\vspace{-3pt}\eq{\fwboxR{0pt}{\fig{-47.65pt}{1}{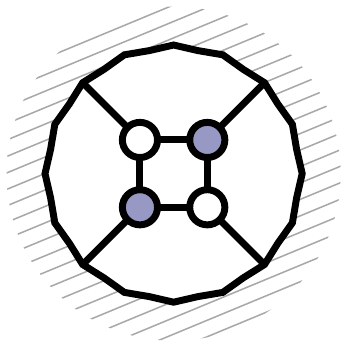}}\bigger{\Leftrightarrow}\fwboxL{0pt}{\hspace{-3.5pt}\fig{-47.65pt}{1}{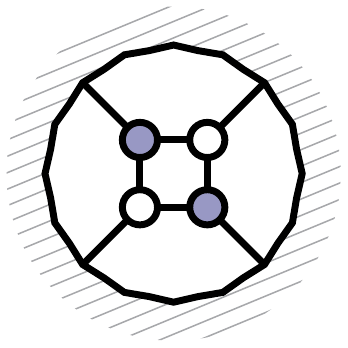}}\label{square_move_identity}\vspace{-3pt}}
This equivalence relation has nothing to do with planarity, and corresponds to a cluster mutation for the on-shell variety. (And because cluster mutations are volume-preserving, this is guaranteed to leave invariant the on-shell function (\ref{grassmannian_representation_of_on_shell_functions}).)

These two equivalence relations leave invariant both on-shell functions and on-shell varieties. For {planar} on-shell diagrams it can be shown that two diagrams correspond to the same positroid variety iff they are related through a sequence of mergers (\ref{merge_expand_identity}) and square moves (\ref{square_move_identity}). It remains an open question whether or not this remains true for non-planar on-shell diagrams. Specifically, is it possible for two on-shell diagrams, not related by mergers and square moves, to correspond to the same on-shell variety or on-shell function? Our surveys have found no example of this happening, and we conjecture that no further equivalence relations among diagrams exist. 

\vspace{-6pt}\subsection{Iteratively Building-Up On-Shell Diagrams and On-Shell Varieties}\label{subsec:building_up_diagrams}\vspace{-6pt}
In the next section, we will review how canonical coordinate charts can be obtained for any on-shell diagram directly. But let us first describe how the correspondence between on-shell diagrams and on-shell varieties can be understood in more geometric terms---by building-up any diagram sequentially from the fundamental vertices (\ref{three_point_vertices}). 

Any on-shell diagram can be constructed through a sequence of two fundamental operations: combining graphs into larger (disconnected) graphs, and gluing legs together. The first of these operations acts in the obvious (and trivial) way on the varieties associated with the graphs. Given on-shell diagrams $\Gamma_{L}$ and $\Gamma_R$, we can define their outer product according to:
\vspace{-6pt}\eq{(\Gamma_L,\Gamma_R)\mapsto(\Gamma_{L}\!\oplus\!\Gamma_{R}),\quad \begin{array}{l}(C_L,C_R)\mapsto (C_{L}\!\oplus\!C_{R})\!\in\!G(k_L\pl k_R,n_L\pl n_R)\\(\Omega_L,\Omega_R)\mapsto\Omega_L\wedge\Omega_R\quad\,\mathrm{with}\,\quad d=d_L\pl d_R\,\end{array}\,.\label{outer_product_definition}\vspace{-4pt}}

Less trivially, any two legs of an on-shell diagram can be glued together, resulting in a diagram with two fewer legs. Notice that this operation reduces $k$ by one (see equation (\ref{k_for_trivalent})). The following illustrates the result of gluing external legs:
\eq{\fwboxR{0pt}{\fig{-42.9pt}{0.75}{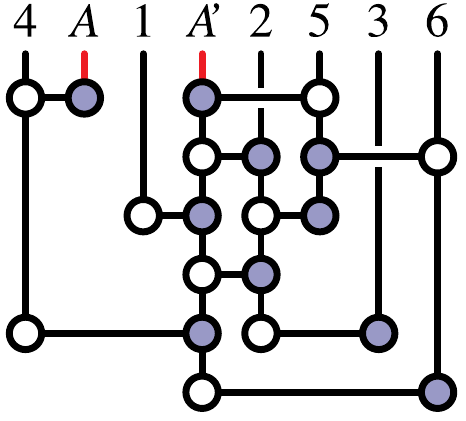}\hspace{6pt}}\bigger{\Rightarrow}\fwboxL{0pt}{\fig{-42.9pt}{0.75}{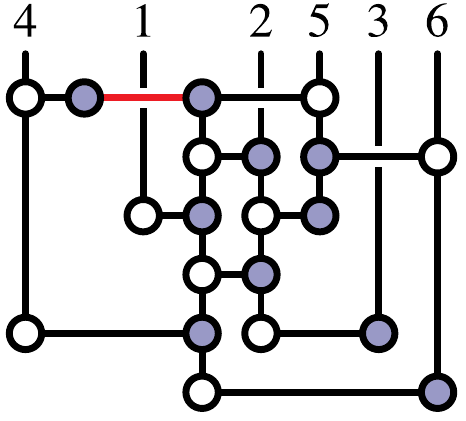}}\label{gluing_illustration}}
This operation on the diagram acts on the corresponding variety according to: 
\eq{\begin{array}{l}G(k\pl1,n\pl2)\ni C_0\mapsto C\!\in\!G(k,n)\quad\mathrm{with}\quad c_{i}\mapsto c_i\cap(c_A\pl c_{A'})^{\perp}\\\Omega_0\equiv\widetilde{\Omega}_0\wedge\frac{d\alpha_A}{\alpha_A}\wedge\frac{d\alpha_{A'}}{\alpha_{A'}}\mapsto\widetilde\Omega_0\wedge\frac{d\alpha_{AA'}}{\alpha_{AA'}}\quad\mathrm{with}\quad d_0\mapsto d_0\mi1\,\end{array}\label{gluing_action_definition}\vspace{-4pt}}
Geometrically, the columns $c_i\!\in\!\mathbb{C}^{k+1}$ of $C_0$ are projected onto the ($k$-dimensional) orthogonal complement of the span of $(c_A\pl c_{A'})$. Notice that the number $d$ of cluster coordinates always reduces by one.

The two operations above are defined without respect to planarity and can be used to iteratively construct the on-shell variety (together with canonical coordinates) from those of the fundamental three-particle vertices (\ref{three_point_vertices}) for any on-shell diagram. Because any non-planar diagram can be constructed from a planar one by iteratively gluing legs together, much of the structure of general on-shell varieties is inherited from that of planar varieties---positroids. While the discussion so far is sufficient to explore general on-shell varieties, it is worth describing one more way that they can be iteratively constructed---through the addition of so-called `BCFW bridges' between legs. While not sufficient to generate all non-planar on-shell diagrams (as evidenced by (\ref{non_bridge_constructible_graphs})), those graphs that can be constructed using BCFW bridges forms an interesting sub-class of on-shell diagrams (including all planar diagrams).

\vspace{-6pt}\paragraph{Building with BCFW Bridges}~\\
\indent One final way that on-shell diagrams can be iteratively built from simpler ones is by adding so-called {\it BCFW bridges}. Adding a bridge `$(a\,{\color{hblue}b})$' attaches a white (blue) vertex to the leg labelled $a$ $({\color{hblue}b})$, and a new internal line between these vertices. The legs $a$ and ${\color{hblue}b}$ need not be adjacent. Under this operation, the on-shell variety is transformed in a very simple way:
\vspace{-4pt}\eq{\begin{array}{c}(a\,{\color{hblue}b})\!:\!(c_1,\ldots,{\color{hblue}c_b},\ldots,c_n)\mapsto (c_1,\ldots,{\color{hblue}c_b'},\ldots,c_n)\quad\mathrm{with}\quad {\color{hblue}c_{b}'}\equiv {\color{hblue}c_b}\pl{\color{hred}\alpha}\,c_a\\\Omega\mapsto\Omega\wedge\frac{d{\color{hred}\alpha}}{{\color{hred}\alpha}}\quad\mathrm{with}\quad d\mapsto d\pl1\,\end{array}\,,\vspace{-4pt}\label{bridge_action_definition}}
shifting column ${{\color{hblue}c_b}}$ by column $c_a$ by an amount parameterized by the (canonical) coordinate ${\color{hred}\alpha}$. Because this new coordinate is canonical, the volume form $\Omega$ is transformed in the obvious way---namely, multiplying it by a factor of $d{\color{hred}\alpha}/{\color{hred}\alpha}$. 

To illustrate this operation, consider the following example:
\vspace{-4pt}\eq{\fwboxR{0pt}{\fig{-31.5pt}{0.8}{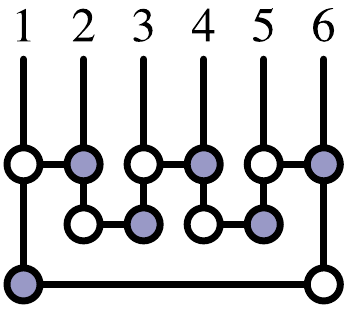}\hspace{7.pt}}\bigger{\underset{\scalebox{0.6}{(1{\color{hblue}3})\hspace{3pt}}}{\Rightarrow}}\fwboxL{0pt}{\fig{-31.5pt}{0.8}{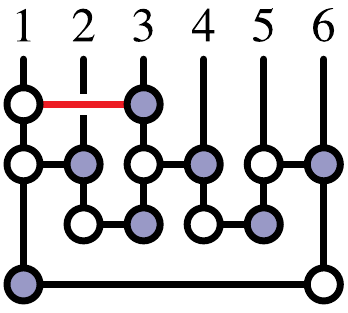}}\label{bcfw_bridge_illustration}}
Starting from any canonical coordinate chart for the initial on-shell variety, we obtain a coordinate chart for the new variety in a simple way. For the example above (\ref{bcfw_bridge_illustration}),
\vspace{-12pt}\eq{\fwboxR{0pt}{\left(\rule{0pt}{25pt}\right.\hspace{-2.5pt}\begin{array}{@{$\,$}cccccc@{}}{\color{black}c_1}&{\color{dim}c_2}&{\color{hblue}c_3}&{\color{dim}c_4}&{\color{dim}c_5}&{\color{dim}c_6}\\\hline{\color{black}0}&{\color{dim}1}&{\color{hblue}\alpha_3}&{\color{dim}\alpha_3\alpha_5}&{\color{dim}0}&{\color{dim}0}\\{\color{black}0}&{\color{dim}0}&{\color{hblue}0}&{\color{dim}1}&{\color{dim}\alpha_2}&{\color{dim}\alpha_2\alpha_4}\\{\color{black}\alpha_1}&{\color{dim}\alpha_1\alpha_6}&{\color{hblue}0}&{\color{dim}0}&{\color{dim}0}&{\color{dim}1}\\~\end{array}\hspace{-4pt}\left.\rule{0pt}{25pt}\right)}\bigger{\underset{\scalebox{0.6}{(1{\color{hblue}3})\hspace{3pt}}}{\Rightarrow}}\fwboxL{0pt}{\left(\rule{0pt}{25pt}\right.\hspace{-2.5pt}\begin{array}{@{$\,$}cccccc@{}}{\color{black}c_1}&{\color{dim}c_2}&{\color{hblue}c_3'}&{\color{dim}c_4}&{\color{dim}c_5}&{\color{dim}c_6}\\\hline{\color{black}0}&{\color{dim}1}&{\color{hblue}\alpha_3}&{\color{dim}\alpha_3\alpha_5}&{\color{dim}0}&{\color{dim}0}\\{\color{black}0}&{\color{dim}0}&{\color{hblue}0}&{\color{dim}1}&{\color{dim}\alpha_2}&{\color{dim}\alpha_2\alpha_4}\\{\color{black}\alpha_1}&{\color{dim}\alpha_1\alpha_6}&\,{\color{hred}\alpha}\,{\color{black}\alpha_1}&{\color{dim}0}&{\color{dim}0}&{\color{dim}1}\\~\end{array}\hspace{-4pt}\left.\rule{0pt}{25pt}\right)\!\!.}\vspace{-12pt}\label{bcfw_bridge_illustration_cmatrices}}

\vspace{0pt}\subsection{(Canonical) Coordinates, Boundary Measurements, and Volume Forms}\label{subsec:canonical_coordinates}\vspace{-6pt}
The discussion above makes it possible to iteratively construct the on-shell variety $C_\Gamma(\alpha)$, parameterized by canonical coordinates, for any on-shell diagram $\Gamma$. But the coordinate chart obtained in this way depends on how the graph is constructed; and no single coordinate chart suffices to expose all of the boundary configurations of a given on-shell variety. The boundaries of an on-shell variety will be described in the next section (where we will also discuss the coordinate charts that allow all boundaries to be reached). But let us first describe a more general approach to constructing the on-shell variety associated with an on-shell diagram. 

Given any on-shell diagram, the variety $C(\alpha)$---represented in terms of canonical coordinates $\alpha_i$---can be obtained as a matrix of {\it boundary measurements} for the graph in the following way. For the sake of this discussion, it is convenient to suppose that the diagram in consideration has been made bipartite (making use of the equivalence (\ref{merge_expand_identity})). 

Boundary measurements for an on-shell diagram are defined with respect to a {\it perfect orientation}---that is, a choice of orientations for the edges of the graph for which every black vertex has a single outgoing edge and every white vertex has a single incoming edge. For a bipartite graph, perfect orientations are in one-to-one correspondence with {\it perfect matchings}---subsets of edges for which every {\it internal} node is an endpoint of exactly one edge in the subset. 

Perfect matchings can be efficiently determined using generalized Kasteleyn matrices, which are certain adjacency matrices for the graph \cite{Franco:2012mm}. The correspondence between perfect matchings and perfect orientations is very simple \cite{2006math09764P,Franco:2012mm}: when a graph is perfectly oriented, there is one preferred edge at every vertex---the one outgoing (incoming) edge at each black (white) vertex; these preferred edges must connect pairs of vertices (from black to white) in non-overlapping subsets, and therefore define a perfect matching. And the construction of a perfect orientation from a perfect matching is similarly straight-forward. The correspondence between perfect orientations and perfect matchings is illustrated in the following example:
\vspace{-6pt}\eq{\fwboxR{0pt}{\fig{-59.85pt}{1}{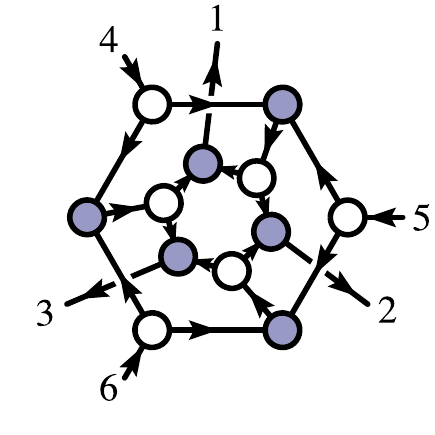}\hspace{10pt}}\bigger{\Leftrightarrow}\fwboxL{0pt}{\hspace{-10pt}\fig{-59.85pt}{1}{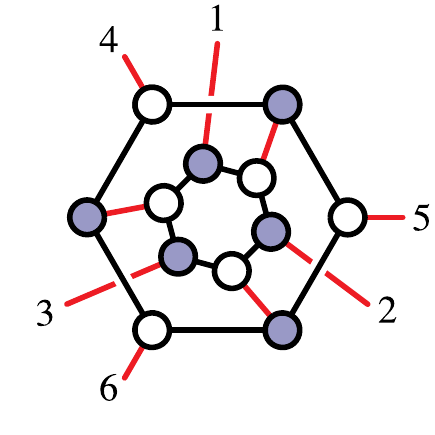}}\label{illustration_of_perfect_matchings_and_prefect_orientations}\vspace{-12pt}}

\newpage
Given an on-shell diagram with a perfect orientation, it easy to construct the matrix $C(\alpha)$ which encodes its boundary measurements. The basic idea is very simple: provided a perfect orientation, an on-shell diagram has $k$ incoming external edges called {\it sources}, and $(n\mi k)$ outgoing edges called {\it sinks}; every edge is assigned a weight $\alpha_e$, and the boundary measurements are the products of edge-weights along connected all paths from the sources to the sinks. See refs.\ \cite{ArkaniHamed:2012nw,Franco:2013nwa} for more details.

To be concrete, suppose the incoming external edges are indexed by $a_i\!\in\!A$ and the sinks indexed by $b\!\in\!B$; then the matrix-representative of the variety $C\!\in\!G(k,n)$ is given by:
\eq{c_b^{i}\equiv \sum_{\gamma\in\{a_i\rightsquigarrow b\}}(-1)^{s_{\gamma}}\prod_{e\in\gamma}\alpha_e\quad \text{with}\quad c^{i}_{a_j}\equiv\delta_{ij}.\label{boundary_measurements_defined}}
Here, the sum runs over all oriented paths $\gamma$ from the source $a_i$ to the sink $b$ (geometrically summing the paths involving closed cycles) and $s_{\gamma}$ represents an important sign that depends on the details of the path $\gamma$. For planar graphs, $s_{\gamma}$ is crucial for the total positivity of the resulting boundary measurement matrix. Although there is no sense of positivity for non-planar graphs, the beautiful combinatorial description of Pl\"{u}cker coordinates in terms of the matroid polytope suggests how these signs should be fixed. These signs were first introduced in \cite{2009arXiv0901.0020G} for diagrams on an annulus, and extended to genus-zero embeddings with an arbitrary number of boundaries in \cite{Franco:2013nwa}; a proposal for generic on-shell diagrams was put forward in \cite{Franco:2015rma}.

For the sake of illustration, the perfect orientation shown in (\ref{illustration_of_perfect_matchings_and_prefect_orientations}) would give rise to the following boundary measurements:
\vspace{-2pt}\eq{\hspace{-35pt}\fwboxR{0pt}{\fig{-47.25pt}{0.8}{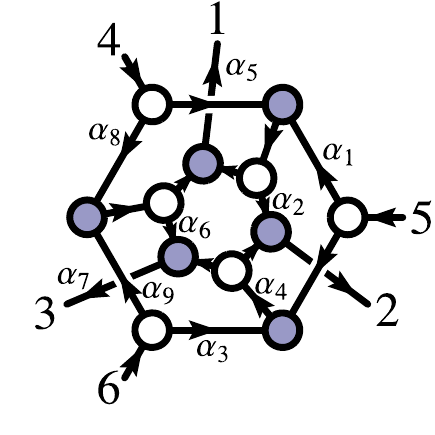}\hspace{1.5pt}}\bigger{\Rightarrow}\,\! C(\alpha)\!\equiv\!\!\left(\rule{0pt}{25pt}\right.\hspace{-4.5pt}\begin{array}{@{}ccc@{}ccc@{}}c_1&c_2&c_3&c_4&c_5&c_6\\\hline\mi\alpha_5(1\pl\alpha_8)&\mi\alpha_2\phantom{\mi}&\alpha_6\,\alpha_7\,\alpha_8&1&{\color{dim}0}&{\color{dim}0}\\\alpha_1\,\alpha_5&\alpha_1\,\alpha_2\pl\alpha_4&\alpha_4\,\alpha_7&{\color{dim}0}&1&{\color{dim}0}\\\mi\alpha_5\,\alpha_9&\alpha_3\,\alpha_4&\alpha_7(\alpha_3\alpha_4\pl\alpha_6\alpha_9)&{\color{dim}0}&{\color{dim}0}&1\\~\end{array}\hspace{-4pt}\left.\rule{0pt}{25pt}\right)\hspace{-100pt}\label{exempli_boundary_measurements}\vspace{-10pt}}
Here, we have set many of the edge-weights to 1 for reasons that we discuss presently.

The attentive reader will notice a discrepancy between the number $d$ of cluster coordinates quoted in equation (\ref{formula_for_d_of_a_graph}) and the number of edges of a graph. This is due to a $GL(1)$-redundancy at each vertex, resulting in $n_V$ extra edge coordinates $\alpha_e$. This means that $n_V$ of the edge weights should be set to 1, explaining why only $9$ of the $21$ edges had non-trivial weights in the example (\ref{exempli_boundary_measurements}). But this redundancy turns out to be useful. Keeping it manifest, the volume form $\Omega$ associated with the boundary measurement matrix $C(\alpha)$ given in (\ref{boundary_measurements_defined}) would be given by,
\eq{\Omega\equiv \left(\prod_{\fwbox{30pt}{\;\;\mathrm{vertices}\, v}} {1\over {\rm vol}(GL(1)_v)}\right)\!\!\left(\prod_{\fwbox{22.5pt}{\;\;{\rm edges }\, e}} {d\alpha_e\over \alpha_e}\right).\label{volume_form_in_edge_variables_with_redundancies}}

\vspace{-6pt}\paragraph{Canonical Coordinates vs. Cluster Coordinates}~\\
\indent For any on-shell variety $C\!\subset\!G(k,n)$ associated with an on-shell diagram, we refer to any coordinate chart as being {\it canonical} if $\Omega$ takes the form given in equation (\ref{volume_form_in_canonical_coordiantes})---a product of factors of the form $d\alpha_i/\alpha_i$. The edge variables associated with boundary measurements appearing in (\ref{volume_form_in_edge_variables_with_redundancies}) are essentially the same as BCFW-bridge coordinates; but while these coordinates are {\it canonical}, they are not strictly speaking {\it cluster} variables: they do not transform under (\ref{square_move_identity}) according to what is ordinarily called a {\it cluster mutation}. However, there does exist closely related coordinates which {\it do} transform as cluster $\mathcal{X}$-coordinates under mutations \cite{ArkaniHamed:2012nw}. And the existence of these coordinates is why on-shell varieties are {\it cluster varieties} \cite{FG2,FG3}.

The cluster coordinates of on-shell varieties are called {\it generalized face variables} and were introduced in ref.\ \cite{Franco:2012wv} (see also refs.\ \cite{Franco:2014nca,Franco:2015rma}). We need not review the details of how these variables are constructed (as their form will not play an important role in our analysis). But the basic idea is to change from edge variables to {face} variables, defined as products of edge-weights along any face of the graph (when embedded on a Riemann surface). The resulting coordinates are still canonical, but with less redundancy than (all) edge variables.

\vspace{-0pt}\paragraph{Pl\"{u}cker Coordinates and Generalized Matroid Data}~\\
\indent Perhaps the most familiar coordinate charts used to describe Grassmannian manifolds are the so-called {\it \Pl coordinates}---which are just the $k\hspace{-0.75pt}\times\hspace{-0.75pt}k$ minors of a matrix representative $C\!\in\!G(k,n)$. As such, \Pl coordinates are labelled by $k$-element subsets of the columns, which we will denote as follows:
\eq{\m{a_1\,a_2\cdots a_k}\equiv\det\!\big\{c_{a_1},c_{a_2},\ldots,c_{a_k}\big\}.\label{definition_of_plucker_coordinate_notation}}
These coordinates satisfy so-called {\it Pl\"{u}cker relations}---which are non-trivial when viewed as abstract coordinates labeled by $k$-element subsets, but follow trivially from Cramer's rule when they are understood as minors. They are also projective: we are free to set any one minor to the identity (corresponding to a choice of $GL(k)$ `gauge' for the matrix representative $C$). For example, \Pl coordinates for $G(3,6)$---in the gauge where minor $(456)$ is set to the identity---would be given by:
\eq{C\!\equiv\!\left(\begin{array}{@{}cccccc@{}}\m{156}&\m{256}&\m{356}&1&{\color{dim}0}&{\color{dim}0}\\\m{416}&\m{426}&\m{436}&{\color{dim}0}&1&{\color{dim}0}\\\m{451}&\m{452}&\m{453}&{\color{dim}0}&{\color{dim}0}&1\end{array}\right)\,.\label{example_of_using_plucker_coordinates}}
Because \Pl coordinates are good coordinates on the Grassmannian, any other coordinate chart can be expressed in these variables. And thus, for the example given in equation (\ref{exempli_boundary_measurements}), it is possible to change variables from $\{\alpha_1,\ldots,\alpha_9\}$ to the \Pl coordinates in (\ref{example_of_using_plucker_coordinates}), and write $\Omega$ in terms of these variables. (The result of this change of variables is given below in equation (\ref{top_form_24_in_plucker_coordinates}).)

Importantly, \Pl coordinates are {\it not canonical} coordinates---and are not obviously the right coordinates in which to write these volume forms. Nevertheless, we will find it useful to express volume-forms of on-shell varieties in these coordinates. One reason for doing this is that the geometric interpretation of the variety is often more clear when the boundaries are viewed as constraints on \Pl coordinates. This is exemplified in the case of positroid varieties, for which the top-form expressed in \Pl coordinates always takes the form,
\eq{\Omega\equiv\Omega_0\times\frac{1}{\m{1\cdots k}\m{2\cdots k\pl1}\cdots\m{n\cdots k\mi1}}\quad\mathrm{with}\quad\Omega_0\equiv\frac{d^{k\times n}C}{\mathrm{vol}(GL(k))}\,.\label{positroid_top_form_in_plucker_coordinates}}
(This form of the volume-form for planar on-shell functions was discovered in \cite{ArkaniHamed:2009dn}.) A more detailed discussion of the meaning of this formula can be found in \mbox{ref.\ \cite{ArkaniHamed:book}}. But we should mention that the factor of `$1/\mathrm{vol}(GL(k))$' appearing in (\ref{positroid_top_form_in_plucker_coordinates}) should be viewed as an instruction to pick a `$GL(k)$-gauge' for the matrix $C$, setting $k$ of the columns to the identity matrix (as in the example (\ref{example_of_using_plucker_coordinates})); and writing the form as `$d^{k\times n}C$' is motivated by the fact that, once such a gauge has been chosen, the non-fixed \Pl coordinates simply correspond to the {\it entries} of the matrix $C\!\equiv\!\big(c^{i}_1,c_2^i,\ldots,c_n^i\big)$. (In the examples discussed below, we will often leave implicit the overall factor $\Omega_0$ when expressing volume forms in \Pl coordinates.)

One advantage to using \Pl coordinates---at least for positroid varieties---is that it makes clear that boundaries correspond to the vanishing of certain \Pl coordinates. Thus, every top-dimensional positroid variety has exactly $n$ co-dimension one sub-strata---obtained by setting one of the consecutive minors in (\ref{positroid_top_form_in_plucker_coordinates}) to zero. This is a consequence of the fact that the cluster coordinates for positroid varieties can always be expressed as ratios of products of \Pl coordinates \cite{ArkaniHamed:book}. This fails to be true more generally, with boundaries corresponding to often intricate relations among \Pl coordinates. 

For example, one of the top-dimensional on-shell varieties of $G(3,6)$---number $9$ according to the classification given in \mbox{Appendix \ref{appendix:top_forms}}---has a volume form that, when expressed in \Pl coordinates, becomes
\eq{\Omega_9\equiv\Omega_0\times\frac{\m{125}}{\m{123}\m{134}\m{156}\m{245}\m{256}\m{16\interBig{25}{34}}}\,,\label{top_form_9_in_plucker_coordinates_exempli}}
where the unusual factor is defined according to:
\eq{\m{16\interBig{25}{34}}\equiv\m{162}\m{534}-\m{165}\m{234}\,.\label{example_of_shifted_column_plucker}}
This notation is motivated by the fact that there exists a unique point in $\mathbb{C}^3$, denoted `$\interBig{25}{34}$', at the intersection of $\mathrm{span}\{c_2,c_5\}$ and $\mathrm{span}\{c_3,c_4\}$. The formula for this point follows easily from (the 3-dimensional form of) Cramer's rule:
\vspace{-4pt}\eq{\interBig{a\,b}{c\,d}\equiv c_a\m{b\,c\,d}-c_{b}\m{a\,c\,d}=-c_c\m{d\,a\,b}+c_d\m{c\,a\,b}\,.\vspace{-4pt}\label{general_form_for_intersections_in_g3n}}
Hence the notation used in the definition (\ref{example_of_shifted_column_plucker}). 

The appearance of such an unusual pole in the volume form $\Omega_9$ in (\ref{top_form_9_in_plucker_coordinates_exempli}) illustrates the richness of geometry that can arise for non-planar on-shell varieties. In particular, this variety has a co-dimension one boundary obtainable by taking the residue on the support of $\m{16\interBig{25}{34}}\!=\!0$---upon which no single \Pl coordinate vanishes. 

Another interesting example is the case of the on-shell diagram and variety given above in equation (\ref{exempli_boundary_measurements}). The volume-form for this variety is denoted $\Omega_{24}$ in our classification; in terms in \Pl coordinates, it is given by:
\vspace{-5pt}\eq{\Omega_{24}\equiv\Omega_0\times\frac{\m{456}^2}{\m{164}\m{145}\m{245}\m{256}\m{356}\m{364}\Delta},\vspace{-5pt}\label{top_form_24_in_plucker_coordinates}}
where the factor $\Delta$ is defined by:
\eq{\Delta\!\equiv\!\sqrt{\big[\m{124}\m{356}\mi\m{234}\m{156}\pl\m{235}\m{146})\big]^2\mi4\,\m{145}\m{356}\m{126}\m{234}}.}
It is not hard to verify that $\Delta$, when evaluated for the boundary measurement matrix $C(\alpha)$ given  in (\ref{exempli_boundary_measurements}), becomes a perfect square. It arises simply as the Jacobian in the transformation between coordinates $\alpha_i$ in (\ref{exempli_boundary_measurements}) to the \Pl coordinates of equation (\ref{example_of_using_plucker_coordinates}).

Interestingly, this on-shell variety has only six boundary configurations (as described in the next section)---all corresponding to the vanishing of one of the \Pl coordinates appearing in the denominator of (\ref{top_form_24_in_plucker_coordinates}). The factor $\Delta$ does not correspond to a boundary (at least not one that corresponds to an on-shell variety); nevertheless, on the support of any of the six \Pl coordinates vanishing, the argument of $\Delta$ becomes a perfect square---and so contributes novel boundaries at co-dimension 2 involving factors similar to that appearing in (\ref{example_of_shifted_column_plucker}).

How to study the intricate structure and geometry of the boundaries of on-shell varieties is described in the next section.

\vspace{-6pt}\subsection{Stratifications of Varieties and Covering Relations}\label{subsec:stratification_and_boundary_relations}\vspace{-6pt}
So far, we have given a rapid sketch of the correspondence between on-shell diagrams and on-shell varieties. In this section, we describe how on-shell varieties are stratified by their covering relations, and how this stratification can be explored even without the combinatorial tools (such as those of \cite{Bourjaily:2012gy}) that exist for positroid varieties. The most important ingredient in this analysis will be the notion of the {\it irreducibility} of on-shell diagrams representing on-shell varieties. Recall that the number $d$ of canonical coordinates for an on-shell variety may exceed the dimension of the corresponding submanifold \mbox{$C(\alpha)\!\subset\!G(k,n)$}. This suggests that we make the following definition. 

\paragraph{Definition:}\label{definition_of_reduced_diagrams} An on-shell diagram is said to be {\it reduced} if the number of canonical coordinates, $d$ (given in equation (\ref{formula_for_d_of_a_graph})), is equal to the dimension of its corresponding on-shell variety $C(\alpha)\!\subset\!G(k,n)$---represented, e.g., by boundary measurements (\ref{boundary_measurements_defined}).\\[-6pt]

An on-shell diagram that is not reduced is said to be {\it reducible}. Because the dimensionality of the on-shell variety is always easy to determine---as the rank of its tangent space, represented in any (possibly degenerate) coordinate chart---the irreducibility of any on-shell diagram may be rapidly determined. Conveniently, the $GL(1)$-redundancies associated with edge variables do not affect this test, and so the boundary-measurement matrix analyzed need not have any edge-weights set to 1. 

Of particular interest are the {\it boundary configurations} of an on-shell variety. These correspond to co-dimension one residues of the volume-form $\Omega$. Recall that this volume-form is constructed precisely so that it has only (and strictly) logarithmic singularities (\ref{volume_form_in_canonical_coordiantes}); and so its co-dimension one residues obviously correspond to the vanishing of some canonical coordinate $\alpha_i$. From the form of $\Omega$ expressed in terms of edge variables in (\ref{volume_form_in_edge_variables_with_redundancies}), these clearly correspond to setting an edge-weight to zero---graphically, to deleting an edge from the on-shell diagram. Thus, boundary configurations can be represented by on-shell diagrams with one edge removed. 

\paragraph{Definition:}\label{definition_of_removable_edge} An edge of an on-shell diagram is said to be {\it removable} if the graph obtained after its deletion is reduced.\\[-6pt]

\noindent Conveniently, this definition allows for removable edges to be identified without reference to any coordinate chart for the corresponding on-shell variety. Thus, given any on-shell diagram it is easy to identify all its removable edges---the removal of which will result in a co-dimension one boundary configuration. For example, consider the on-shell diagram representing variety numbered $20$ in the classification given in \mbox{Appendix \ref{appendix:top_forms}}; for this diagram, we can easily identify its three removable edges, drawn in red below:\\[-12pt]
\vspace{-0pt}\eq{\fig{-46pt}{0.8}{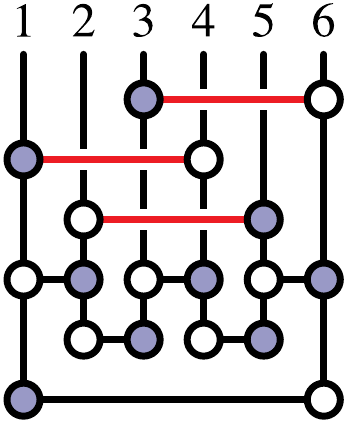}\label{top_form_20_with_removable_edges_indicated}\vspace{-0pt}}

For planar on-shell diagrams, there is a one-to-one correspondence between removable edges and boundary configurations. This, however, is not true in general for non-planar diagrams: the correspondence can be many-to-one. While the varieties obtained from diagrams where removable edges have been removed surely correspond to boundary configurations, it is not generally true that they are all distinct. 

Among the best illustrations of this new phenomenon is the first on-shell diagram drawn in equation (\ref{non_bridge_constructible_graphs}). This diagram has twelve removable edges but only six boundaries; its removable edges correspond to those highlighted,
\vspace{-6pt}\eq{\fig{-47.25pt}{0.8}{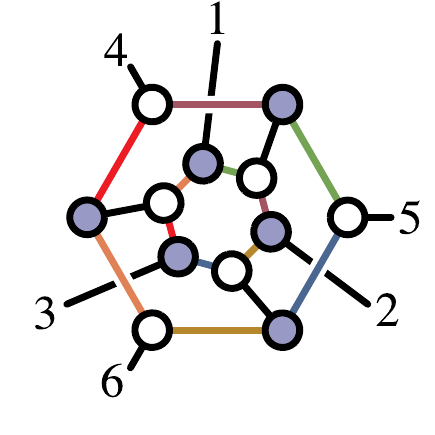}\label{removable_edges_of_diagram_24_highlighted_in_pairs}\vspace{-10pt}}
Here, we have colored each removable edge according to the boundary configuration that results. Removable edges drawn in the same color correspond to {\it identical} boundaries. Consider for example the pair of edges drawn in red. Removing each edge results in a different reduced on-shell diagram:
\vspace{-8pt}\eq{\fwboxR{0pt}{\fig{-47.25pt}{0.8}{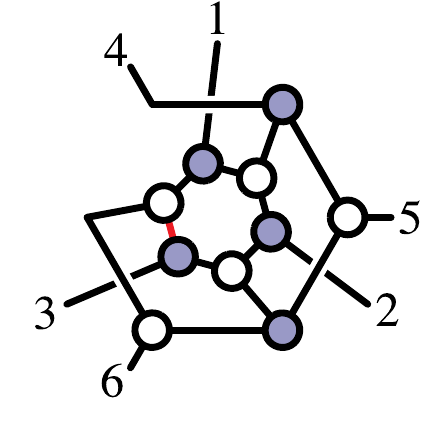}\hspace{10pt}}\bigger{\sim}\fwboxL{0pt}{\fig{-47.25pt}{0.8}{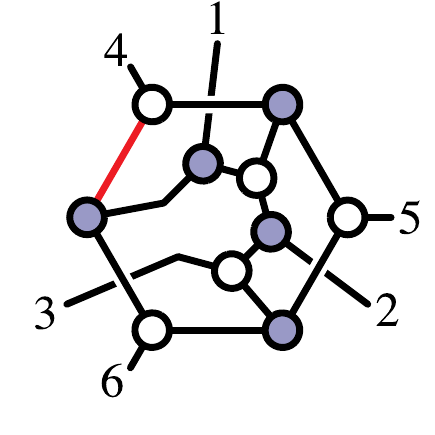}}\label{equivalent_boundaries_fig}\vspace{-15pt}}
In this case, the equivalence between the two varieties can be confirmed by performing a square move on each graph (the only one possible in each case). It is worth mentioning that there does not exist any ($GL(1)$-reduced) canonical edge-coordinate chart for which both edges have non-unit weight. 

The fact that the removal of {\it distinct} edges of an on-shell diagram can result in the {\it same} boundary variety is a troubling novelty of non-planar diagrams. It is troubling because, without explicitly constructing the mutations required to show that two boundary elements are the same, a more systematic approach to the identification of equivalent on-shell varieties is required. Let us be clear about the meaning of identical varieties, mentioned in passing above. 

\paragraph{Definition:}\label{definition_of_identical_varieties} Two on-shell varieties are said to be {\it identical} if there exists a non-singular change of variables (with unit Jacobian) between their canonical coordinates.\\[-8pt]

Two varieties that are not identical are called {\it distinct}. The varieties of diagrams related by square moves and mergers are clearly identical, but it is unknown whether all diagrams corresponding to identical varieties are related by sequences of these equivalence transformations. Without knowing an explicit coordinate transformation, how can identical varieties be identified? We propose the following test:

\paragraph{Conjecture:}\label{equivalence_of_poset_conjecture} The on-shell varieties of two on-shell diagrams are {\it identical} iff their collection of boundary configurations are identical.\\[-6pt]

\noindent Because at sufficiently low dimension all on-shell varieties are positroid varieties for which this is trivially true, this conjecture implies a systematic, recursive test for identifying identical configurations. And so provided this, we may define:

\paragraph{Definition:}\label{definition_of_stratifications} We will call the graph generated by covering relations---connecting a diagram to its distinct boundaries---for a given on-shell diagram its {\it stratification}.\\[-6pt]

Because permutations of leg labels, and parity (exchanging the colors of vertices throughout) clearly leave an on-shell variety's stratification invariant as a graph, this data is well suited for distinguishing fundamentally inequivalent varieties. 

When two varieties are related by a reordering of the legs or parity, we call them {\it equivalent}. ({\it Nota bene:} we are using `equivalent' and `identical' quite differently!) And in order to identify all equivalence classes of on-shell varieties, we make use of one final conjecture:

\paragraph{Conjecture:}\label{equivalence_of_strata_conjecture} If the stratifications of two on-shell diagrams are isomorphic as graphs, then their corresponding varieties are equivalent (by relabeling legs and/or parity).\\[-6pt]

\noindent Our systematic but incomplete study of examples in $G(3,6)$ suggest that this conjecture is true. However, it amounts to an important caveat regarding the completeness of our classification of on-shell varieties for $G(3,6)$: if either of the two conjectures above fail, there may exist new, inequivalent on-shell varieties missed by our analysis. It is really the space of {\it stratifications} that we have classified for $G(3,6)$. 

Let us conclude this section with an important, exceptional novelty discovered for $G(3,6)$. For the top-form $\Omega_{24}$ given in (\ref{top_form_24_in_plucker_coordinates}) corresponding to the variety (\ref{exempli_boundary_measurements}), we have found that the sum of {\it inequivalent} boundaries associated with removable edges (shown in (\ref{removable_edges_of_diagram_24_highlighted_in_pairs})) does not correspond to a residue theorem. This suggests that there may be residues associated with taking some edge variables to infinity---which would not correspond to an on-shell diagram (at least not one obtained from deleting edges). Such behavior has never before been seen in $\mathcal{N}\!=\!4$ SYM, although it is prevalent in $\mathcal{N}\!=\!8$ supergravity, \cite{Herrmann:2016qea,Heslop:2016plj}. This may be related to poles at infinity, which were expected to be absent in $\mathcal{N}\!=\!4$ SYM even beyond the planar limit, \cite{Arkani-Hamed:2014via}. We do not know what the meaning or implications of this novelty, but it does not affect our analysis of on-shell varieties (defined to be those related to on-shell diagrams). More likely, however, we think that the correct prescription for the residue theorem involves all boundary configurations obtained by removing removable edges, including equivalent copies with multiplicity. This renders the residue theorem trivial, as the 12 removable edges of $\Omega_{24}$ come in 6 equivalent pairs.

\vspace{-6pt}\section{Stratifying the Varieties of (General) On-Shell Diagrams}\label{sec:stratification_of_top_forms}\vspace{-6pt}
We are interested in classifying all cluster varieties associated with on-shell diagrams---considering those related by relabeling the external legs of the diagram to be equivalent. Relabeling the external legs corresponds to a permutation of the columns of the matrix (representative) $C\!\equiv\!\big(c_1,c_2,\ldots, c_n)\!\in\!G(k,n)$ of the on-shell variety. We will also consider varieties related by parity, $C\!\mapsto\!C^\perp\!\in\!G(n\mi k,n)$, to be equivalent (relevant when $n\!=\!2k$). The on-shell varieties of planar on-shell diagrams are called positroid varieties, which are completely characterized by (decorated) permutations as described in \mbox{ref.\ \cite{ArkaniHamed:book}}. But as mentioned above, little is known about the scope of varieties that exist beyond the planar limit. 

A systematic way to explore the space of on-shell varieties beyond the planar limit is to simply construct---by brute force---all reduced on-shell diagrams for fixed $k,n$ and identify the equivalence classes that result. We describe below how the space of non-planar diagrams can be explored exhaustively (for fixed $k,n$). Let us now describe how all on-shell diagrams can be exhaustively (and exhaustingly) constructed.

\vspace{-6pt}\subsection{Constructing Representatives of All On-Shell Diagrams}\label{subsec:generating_all_diagrams}\vspace{-6pt}
Non-planar on-shell diagrams can be constructed from planar ones by gluing together general pairs of legs, or by attaching BCFW bridges between non-adjacent legs. Neither of these operations depends on planarity. While the first is clearly the only general strategy, it turns out to be much easier to enumerate the graphs generated by sequences of BCFW bridges. Thus, we find it advantageous to separately enumerate the on-shell diagrams that are `bridge-constructible'---those obtainable by sequences of (possibly non-adjacent) bridges---and then those that are non-bridge constructible.

\subsubsection{Bottom-Up Approach: Bridge-Constructible Diagrams}\label{subsubsec:bridge_construction}\vspace{-6pt}
Let us first describe the enumeration of all bridge-constructible on-shell varieties. These correspond to varieties represented by diagrams constructible from an empty diagram through a sequence of (possibly non-adjacent) BCFW bridges. All positroid varieties are bridge constructible; but this turns out not to be true for non-planar varieties, as we have seen in (\ref{non_bridge_constructible_graphs}). Nevertheless, it is computationally much easier to enumerate all sequences of BCFW bridges, providing an interesting (though incomplete) subset of general on-shell varieties. 

Consider any bridge constructible on-shell variety for the top-cell of $G(k,n)$. This can be represented by a reduced diagram with $(k\mi1)(n\mi k\mi1)$ internal cycles. Removing a BCFW bridge results in a diagram with one fewer internal cycles. And so after enough boundaries have been taken, the resulting diagram is guaranteed to be planar with respect to some ordering of the external legs. Therefore, we need not consider {\it all} diagrams constructed through sequences of bridges; rather, we need only consider those obtainable through sequences of bridges starting from representatives of inequivalent low-dimensional positroid varietys. This greatly simplifies the analysis, allowing for a rigorously complete list of representatives diagrams to be constructed.

\vspace{-6pt}\subsubsection{Top-Down Approach: Non-Planar Diagrams from Gluing Legs}\label{subsubsec:gluing_construction}\vspace{-6pt}
Consider an arbitrary on-shell diagram, representing an on-shell variety in $G(k,n)$. By cutting open an internal line, it can always be obtained from a diagram with two additional legs, representing a variety in $G(k\pl1,n\pl2)$. Because cutting open an internal line results in a graph with one fewer internal cycle, it is clear that after cutting open a sufficient number, $\ell$, of internal lines the result will be a planar graph in $G(k\pl\ell,n\pl2\ell)$. 

Starting from a top-dimensional variety of $G(k,n)$, it is not hard to see that when $\ell\!=\!k(n\mi k)\mi n$ internal lines are cut, the result is guaranteed to be planar and hence correspond to a positroid variety. For $G(3,6)$, this means that all on-shell diagrams can be constructed from planar diagrams in $G(6,12)$. For example, the first (non-bridge-constructible) diagram in (\ref{non_bridge_constructible_graphs}) can be constructed as follows:
\vspace{-8pt}\eq{\fwboxR{0pt}{\fig{-47.25pt}{0.8}{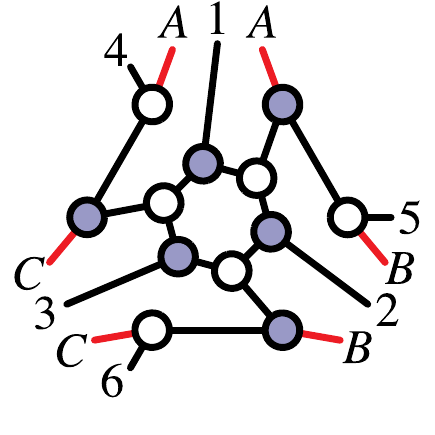}\hspace{10pt}}\bigger{\Rightarrow}\fwboxL{0pt}{\hspace{0pt}\fig{-47.25pt}{0.8}{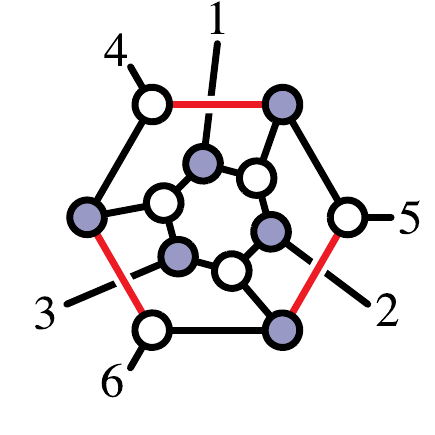}}}
~\\[-20pt]
Similarly, the second (non-bridge-constructible) diagram in (\ref{non_bridge_constructible_graphs}) can be constructed from a diagram in $G(5,10)$ as follows:
\vspace{-8pt}\eq{\fwboxR{0pt}{\fig{-32pt}{0.8}{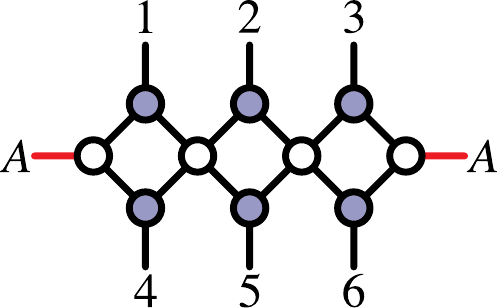}\hspace{7.5pt}}\hspace{0pt}\bigger{\Rightarrow}\hspace{0pt}\fwboxL{0pt}{\fig{-41.1pt}{0.8}{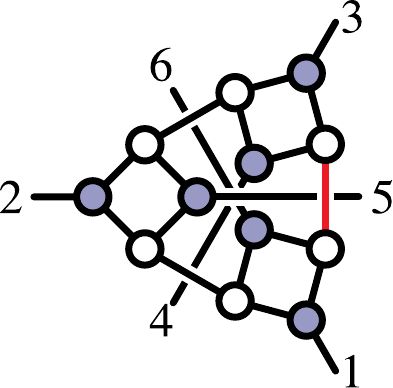}}}
~\\[-12pt]
\indent For the purposes of classification, it is worth bearing in mind that we need only consider those on-shell diagrams which, after gluing, are non-bridge constructible. Moreover, cutting upon an internal line can always result in a graph with at least one bridge-removable boundary. (For $G(3,6)$, it can be further shown that any non-bridge-constructible graph is fully bridge-constructible after one internal line has been cut.) This greatly reduces the space of possible starting points, allowing for a systematic survey to be completed.

\vspace{-0pt}\section{The Stratification of On-Shell Varieties for $G(3,6)$}\label{sec:g36_classification}\vspace{-6pt}
In the case of $G(3,6)$, top-dimensional on-shell varieties correspond to reduced on-shell diagrams with 4 internal cycles. Thus, a complete survey of bridge-constructible graphs can be obtained by starting with representative diagrams for all $6$-dimensional positroid varieties of $G(3,6)$, and all non-bridge-constructible graphs obtained by iteratively gluing pairs of legs of (12-dimensional) positroid varieties of $G(6,12)$. 

Among this large, but manageable list of on-shell diagrams, only 24 inequivalent on-shell varieties were found. Of these, 22 are directly bridge-constructible (and, interestingly, all constructed as sequence of bridges starting from a single same 6-dimensional positroid variety). These are enumerated in \mbox{Appendix \ref{appendix:top_forms}}, where we have also written their volume-forms in terms of Pl\"ucker coordinates. Details for this classification are also provided in the file {\tt g36\_top\_form\_data.txt}, included as part of this work's submission files on the {\tt arXiv}. 

From these 24 top-dimensional varieties, all lower-dimensional on-shell varieties can be obtained by taking boundaries---by removing removable edges. At co-dimension one, we find only 10 inequivalent on-shell varieties, representatives of which are given in \mbox{Appendix \ref{appendix:leading_singularities}}. Being related to so-called `leading singularities', these are of particular interest to physics. And so, we have also provided explicit, spinor-helicity formulae for each of these. 

Because relations among on-shell functions can be obtained as residue theorems starting from a higher-dimensional variety, in \mbox{Appendix \ref{appendix:leading_singularities}} we give explicit formulae for all the residue theorems generated by these varieties. The exceptional case described above for the boundaries of $\Omega_{24}$ is included among this list, although not an identity among (the on-shell functions of) on-shell varieties. 

In \mbox{Appendix \ref{appendix:higher_singularities}} we continue this classification to lower dimensions. Representative diagrams are given for all inequivalent classes of 7- and 6-dimensional on-shell varieties---those with lower dimension are all positroids. The enumeration of different varieties for $G(3,6)$ is summarized in \mbox{Table \ref{table_of_strata}}. Here, we have also listed the numbers of {\it prime} on-shell varieties, defined to be those which are not the disconnected outer-product of non-trivial on-shell varieties (allowing for arbitrary numbers of disconnected, zero-dimensional components).\\[-20pt]

\begin{table}[b]\vspace{-40pt}$$\fwbox{430pt}{\begin{array}{|r|c|c|c|c|c|c|c|c|c|c|}\cline{2-11}\multicolumn{1}{r|}{\text{dim:}}&9&8&7&6&5&4&3&2&1&0\\\hline\text{prime:}&24&10&7&6&4&3&2&1&1&1\\\hline\text{planar:}&1&1&3&5&5&5&4&2&1&1\\\hline\text{total:}&24&10&7&6&5&5&4&2&1&1\\\hline\end{array}}$$\vspace{-20pt}\caption{The number of inequivalent on-shell varieties of $G(3,6)$ for each dimension. Recall that a {\it prime} variety is one which is not a product manifold (but may involve zero-dimensional, disconnected components consisting of `hanging' legs).}\vspace{-20pt}\label{table_of_strata}\end{table}

\newpage
\section{Conclusions and Future Directions}\label{sec:conclusions}\vspace{-6pt}
Non-planarity represents one of the obvious future frontiers in the study of on-shell diagrams. This problem is interesting both for its potential applications to scattering amplitudes and for its new mathematical and geometric structures. 

In this paper we have put forth a strategy toward the classification of inequivalent on-shell varieties in $G(k,n)$ corresponding to generally non-planar on-shell diagrams. The classification of planar on-shell diagrams has lead to beautiful geometric and combinatorial tools. These tools become insufficient when abandoning planarity. We advocate the stratification of on-shell diagrams, for which we developed efficient tools, as a way of identifying equivalence classes. We applied our approach to $G(3,6)$, identifying 24 inequivalent top-dimensional on-shell varieties.

Our investigation reveals that non-planar on-shell diagrams give rise to a variety of new phenomena that include: a number of codimension-one boundaries that can differ from the total number of \Pl coordinates in the denominator, poles at which no (single) \Pl coordinate vanishes, boundary operator that does not square to zero, multiple ways of accessing a boundary from a diagram one dimension above, the appearance of square roots in the volume form when expressing it in term of \Pl coordinates, and the signal of possible poles at infinity. These novelties suggest that much further work needs to be done to understand the systematics of what is possible for the on-shell varieties beyond the case of $G(3,6)$. 

There are various clear directions for future research. It would be interesting to study the translation between the on-shell forms that result from our classification and configurations of points in momentum twistor space. Another natural question is whether the combination of non-equivalent on-shell varieties define some interesting region in $G(k,n)$ or whether it tiles it completely. Finally, it would be very interesting to understand the infinite dimensional symmetries of non-planar on-shell functions that generalize the Yangian (along the lines of \cite{Frassek:2016wlg}). The answers to these and similar questions will likely shed light on the application of on-shell diagrams to computation of non-planar scattering amplitudes.

\section*{Acknowledgements}\vspace{-6pt}
We are grateful for helpful discussions with Jaroslav Trnka, Alexander Postnikov, and Lauren Williams, and for the hospitality of the University of California, Davis. This work was funded in part by the Danish National Research Foundation (DNRF91) and the Danish Council for Independent Research (JLB); and by the US National Science Foundation under grant PHY-1518967 and a PSC-CUNY award (SF).

\newpage
\appendix\section{Representative On-Shell Diagrams for Top-Forms of $G(3,6)$}\label{appendix:top_forms}\vspace{-12pt}
We have found 24 permutation (and parity) inequivalent, top-dimensional on-shell varieties for $G(3,6)$. These forms, labeled $\Omega_{i}$, are given below together with the stratification numbers and representative on-shell diagrams for each. Machine-readable details for each archetype are provided as part of this work's submission files to the {\tt arXiv}, in the file {\tt g36\_top\_form\_data.txt}.
\vspace{-2pt}\eq{\fwbox{430pt}{\begin{array}{|@{}l@{}|@{}l@{}|@{}r@{}|}\hline\fwboxL{25pt}{\,\,\Omega_{1}}&\fwboxL{0pt}{\text{   strat.\ numbers: }}\fwboxL{280pt}{\hspace{85pt}\{1,\!6,\!21,\!56,\!114,\!180,\!215,\!180,\!90,\!20\}}&\fwbox{50pt}{\chi=1}\\\hline\multicolumn{3}{|@{}l@{}|}{\fwbox{0pt}{\rule{0.pt}{40.75pt}}\raisebox{16.15pt}{$\fwboxL{355pt}{\hspace{5pt}\displaystyle\scalebox{0.85}{$\displaystyle\frac{1}{\m{123}\m{234}\m{345}\m{456}\m{561}\m{612}}$}}$}}\\\hline\end{array}\,\begin{array}{@{}c@{}}\\[-20pt]\hspace{-3.95pt}\fig{-42.9pt}{0.75}{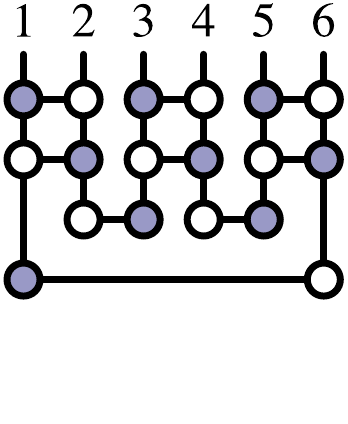}\\[-32pt]\end{array}\nonumber}}
\vspace{-2pt}\eq{\fwbox{430pt}{\begin{array}{|@{}l@{}|@{}l@{}|@{}r@{}|}\hline\fwboxL{25pt}{\,\,\Omega_{2}}&\fwboxL{0pt}{\text{   strat.\ numbers: }}\fwboxL{280pt}{\hspace{85pt}\{1,\!7,\!27,\!83,\!166,\!239,\!249,\!190,\!90,\!20\}}&\fwbox{50pt}{\chi=6}\\\hline\multicolumn{3}{|@{}l@{}|}{\fwbox{0pt}{\rule{0.pt}{70.75pt}}\raisebox{36.15pt}{$\fwboxL{355pt}{\hspace{5pt}\displaystyle\scalebox{0.85}{$\displaystyle\frac{\m{235}}{\m{123}\m{136}\m{156}\m{234}\m{245}\m{256}\m{345}}$}}$}}\\\hline\end{array}\,\begin{array}{@{}c@{}}\hspace{-3.95pt}\fig{-42.9pt}{0.75}{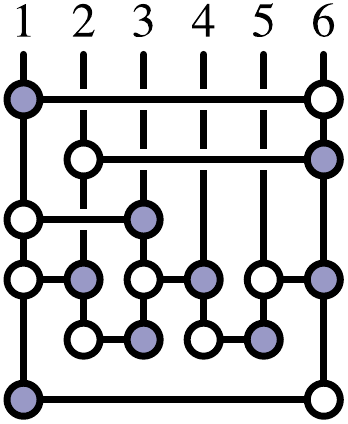}\end{array}\nonumber}}
\vspace{-2pt}\eq{\fwbox{430pt}{\begin{array}{|@{}l@{}|@{}l@{}|@{}r@{}|}\hline\fwboxL{25pt}{\,\,\Omega_{3}}&\fwboxL{0pt}{\text{   strat.\ numbers: }}\fwboxL{280pt}{\hspace{85pt}\{1,\!8,\!30,\!98,\!198,\!274,\!268,\!195,\!90,\!20\}}&\fwbox{50pt}{\chi=8}\\\hline\multicolumn{3}{|@{}l@{}|}{\fwbox{0pt}{\rule{0.pt}{70.75pt}}\raisebox{36.15pt}{$\fwboxL{355pt}{\hspace{5pt}\displaystyle\scalebox{0.85}{$\displaystyle\frac{\m{235}^2}{\m{123}\m{135}\m{156}\m{234}\m{236}\m{245}\m{256}\m{345}}$}}$}}\\\hline\end{array}\,\begin{array}{@{}c@{}}\hspace{-3.95pt}\fig{-42.9pt}{0.75}{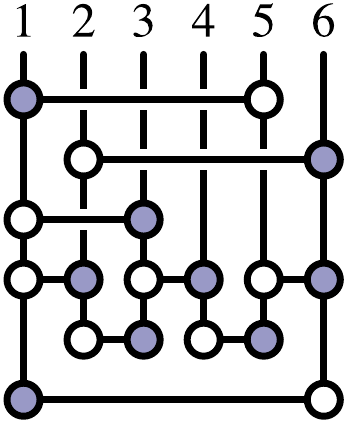}\end{array}\nonumber}}
\vspace{-2pt}\eq{\fwbox{430pt}{\begin{array}{|@{}l@{}|@{}l@{}|@{}r@{}|}\hline\fwboxL{25pt}{\,\,\Omega_{4}}&\fwboxL{0pt}{\text{   strat.\ numbers: }}\fwboxL{280pt}{\hspace{85pt}\{1,\!8,\!34,\!116,\!215,\!282,\!271,\!196,\!90,\!20\}}&\fwbox{50pt}{\chi=11}\\\hline\multicolumn{3}{|@{}l@{}|}{\fwbox{0pt}{\rule{0.pt}{55.75pt}}\raisebox{26.15pt}{$\fwboxL{355pt}{\hspace{5pt}\displaystyle\scalebox{0.85}{$\displaystyle\frac{\m{135}\m{145}}{\m{123}\m{125}\m{134}\m{136}\m{156}\m{245}\m{345}\m{456}}$}}$}}\\\hline\multicolumn{3}{@{}c@{}}{\rule{0pt}{10pt}}\\[-16pt]\end{array}\,\begin{array}{@{}c@{}}\\[-20pt]\hspace{-3.95pt}\fig{-42.9pt}{0.75}{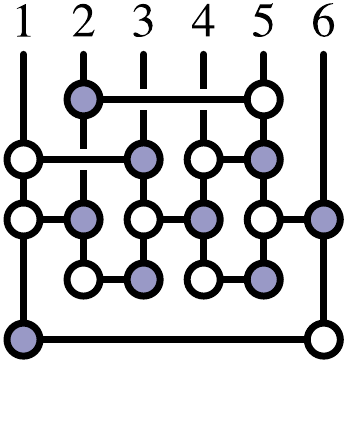}\\[-20pt]\end{array}\vspace{-50pt}\nonumber}}
\vspace{-2pt}\eq{\fwbox{430pt}{\begin{array}{|@{}l@{}|@{}l@{}|@{}r@{}|}\hline\fwboxL{25pt}{\,\,\Omega_{5}}&\fwboxL{0pt}{\text{   strat.\ numbers: }}\fwboxL{280pt}{\hspace{85pt}\{1,\!9,\!36,\!138,\!252,\!315,\!288,\!201,\!90,\!20\}}&\fwbox{50pt}{\chi=16}\\\hline\multicolumn{3}{|@{}l@{}|}{\fwbox{0pt}{\rule{0.pt}{70.75pt}}\raisebox{36.15pt}{$\fwboxL{355pt}{\hspace{5pt}\displaystyle\scalebox{0.85}{$\displaystyle\frac{\m{135}^3}{\m{123}\m{125}\m{134}\m{136}\m{145}\m{156}\m{235}\m{345}\m{356}}$}}$}}\\\hline\end{array}\,\begin{array}{@{}c@{}}\hspace{-3.95pt}\fig{-42.9pt}{0.75}{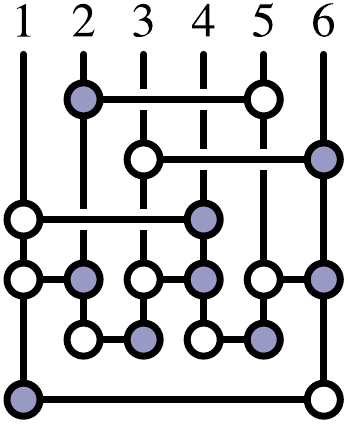}\end{array}\nonumber}}
\vspace{-2pt}\eq{\fwbox{430pt}{\begin{array}{|@{}l@{}|@{}l@{}|@{}r@{}|}\hline\fwboxL{25pt}{\,\,\Omega_{6}}&\fwboxL{0pt}{\text{   strat.\ numbers: }}\fwboxL{280pt}{\hspace{85pt}\{1,\!9,\!38,\!122,\!236,\!309,\!285,\!199,\!90,\!20\}}&\fwbox{50pt}{\chi=9}\\\hline\multicolumn{3}{|@{}l@{}|}{\fwbox{0pt}{\rule{0.pt}{55.75pt}}\raisebox{26.15pt}{$\fwboxL{355pt}{\hspace{5pt}\displaystyle\scalebox{0.85}{$\displaystyle\frac{\m{145}\m{235}^2}{\m{123}\m{125}\m{135}\m{156}\m{234}\m{236}\m{245}\m{345}\m{456}}$}}$}}\\\hline\multicolumn{3}{@{}c@{}}{\rule{0pt}{10pt}}\\[-16pt]\end{array}\,\begin{array}{@{}c@{}}\\[-20pt]\hspace{-3.95pt}\fig{-42.9pt}{0.75}{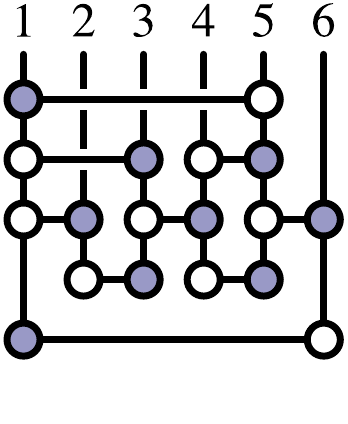}\\[-20pt]\end{array}\vspace{-50pt}\nonumber}}
\vspace{-2pt}\eq{\fwbox{430pt}{\begin{array}{|@{}l@{}|@{}l@{}|@{}r@{}|}\hline\fwboxL{25pt}{\,\,\Omega_{7}}&\fwboxL{0pt}{\text{   strat.\ numbers: }}\fwboxL{280pt}{\hspace{85pt}\{1,\!8,\!36,\!102,\!189,\!256,\!257,\!192,\!90,\!20\}}&\fwbox{50pt}{\chi=5}\\\hline\multicolumn{3}{|@{}l@{}|}{\fwbox{0pt}{\rule{0.pt}{55.75pt}}\raisebox{26.15pt}{$\fwboxL{355pt}{\hspace{5pt}\displaystyle\scalebox{0.85}{$\displaystyle\frac{\m{16\inter{23}{45}}}{\m{123}\m{126}\m{136}\m{156}\m{234}\m{245}\m{345}\m{456}}$}}$}}\\\hline\multicolumn{3}{@{}c@{}}{\rule{0pt}{10pt}}\\[-16pt]\end{array}\,\begin{array}{@{}c@{}}\\[-20pt]\hspace{-3.95pt}\fig{-42.9pt}{0.75}{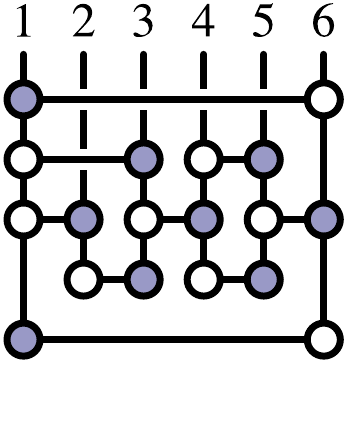}\\[-20pt]\end{array}\vspace{-50pt}\nonumber}}
\vspace{-2pt}\eq{\fwbox{430pt}{\begin{array}{|@{}l@{}|@{}l@{}|@{}r@{}|}\hline\fwboxL{25pt}{\,\,\Omega_{8}}&\fwboxL{0pt}{\text{   strat.\ numbers: }}\fwboxL{280pt}{\hspace{85pt}\{1,\!10,\!45,\!142,\!267,\!334,\!297,\!202,\!90,\!20\}}&\fwbox{50pt}{\chi=8}\\\hline\multicolumn{3}{|@{}l@{}|}{\fwbox{0pt}{\rule{0.pt}{55.75pt}}\raisebox{26.15pt}{$\fwboxL{355pt}{\hspace{5pt}\displaystyle\scalebox{0.85}{$\displaystyle\frac{\m{16\inter{23}{45}}^2}{\m{123}\m{126}\m{136}\m{146}\m{156}\m{234}\m{235}\m{245}\m{345}\m{456}}$}}$}}\\\hline\multicolumn{3}{@{}c@{}}{\rule{0pt}{10pt}}\\[-16pt]\end{array}\,\begin{array}{@{}c@{}}\\[-20pt]\hspace{-3.95pt}\fig{-42.9pt}{0.75}{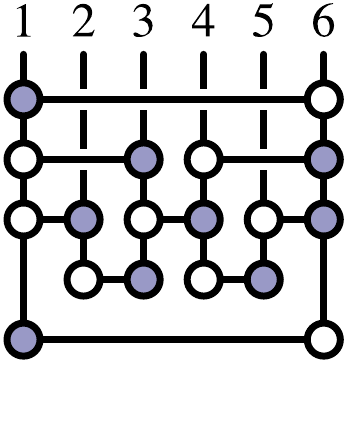}\\[-20pt]\end{array}\vspace{-50pt}\nonumber}}
\vspace{-2pt}\eq{\fwbox{430pt}{\begin{array}{|@{}l@{}|@{}l@{}|@{}r@{}|}\hline\fwboxL{25pt}{\,\,\Omega_{9}}&\fwboxL{0pt}{\text{   strat.\ numbers: }}\fwboxL{280pt}{\hspace{85pt}\{1,\!6,\!25,\!78,\!158,\!231,\!245,\!189,\!90,\!20\}}&\fwbox{50pt}{\chi=5}\\\hline\multicolumn{3}{|@{}l@{}|}{\fwbox{0pt}{\rule{0.pt}{70.75pt}}\raisebox{36.15pt}{$\fwboxL{355pt}{\hspace{5pt}\displaystyle\scalebox{0.85}{$\displaystyle\frac{\m{125}}{\m{123}\m{134}\m{156}\m{245}\m{256}\m{16\inter{25}{34}}}$}}$}}\\\hline\end{array}\,\begin{array}{@{}c@{}}\hspace{-3.95pt}\fig{-42.9pt}{0.75}{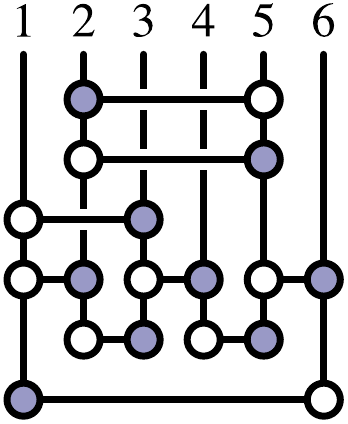}\end{array}\nonumber}}
\vspace{-2pt}\eq{\fwbox{430pt}{\begin{array}{|@{}l@{}|@{}l@{}|@{}r@{}|}\hline\fwboxL{25pt}{\,\,\Omega_{10}}&\fwboxL{0pt}{\text{   strat.\ numbers: }}\fwboxL{280pt}{\hspace{85pt}\{1,\!7,\!29,\!107,\!209,\!280,\!271,\!196,\!90,\!20\}}&\fwbox{50pt}{\chi=10}\\\hline\multicolumn{3}{|@{}l@{}|}{\fwbox{0pt}{\rule{0.pt}{70.75pt}}\raisebox{36.15pt}{$\fwboxL{355pt}{\hspace{5pt}\displaystyle\scalebox{0.85}{$\displaystyle\frac{\m{234}\m{235}}{\m{123}\m{134}\m{236}\m{245}\m{256}\m{345}\m{14\inter{23}{56}}}$}}$}}\\\hline\end{array}\,\begin{array}{@{}c@{}}\hspace{-3.95pt}\fig{-42.9pt}{0.75}{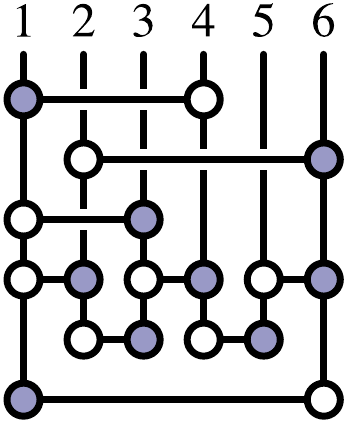}\end{array}\nonumber}}
\vspace{-2pt}\eq{\fwbox{430pt}{\begin{array}{|@{}l@{}|@{}l@{}|@{}r@{}|}\hline\fwboxL{25pt}{\,\,\Omega_{11}}&\fwboxL{0pt}{\text{   strat.\ numbers: }}\fwboxL{280pt}{\hspace{85pt}\{1,\!7,\!33,\!104,\!194,\!261,\!260,\!193,\!90,\!20\}}&\fwbox{50pt}{\chi=7}\\\hline\multicolumn{3}{|@{}l@{}|}{\fwbox{0pt}{\rule{0.pt}{70.75pt}}\raisebox{36.15pt}{$\fwboxL{355pt}{\hspace{5pt}\displaystyle\scalebox{0.85}{$\displaystyle\frac{\m{126}\m{235}}{\m{123}\m{136}\m{156}\m{234}\m{245}\m{256}\m{16\inter{25}{34}}}$}}$}}\\\hline\end{array}\,\begin{array}{@{}c@{}}\hspace{-3.95pt}\fig{-42.9pt}{0.75}{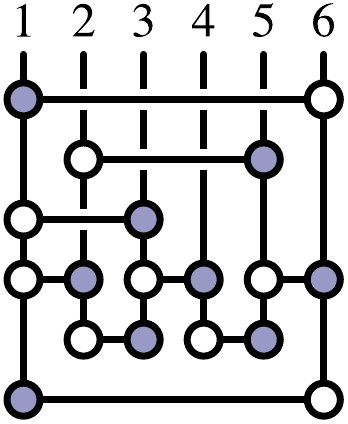}\end{array}\nonumber}}
\vspace{-2pt}\eq{\fwbox{430pt}{\begin{array}{|@{}l@{}|@{}l@{}|@{}r@{}|}\hline\fwboxL{25pt}{\,\,\Omega_{12}}&\fwboxL{0pt}{\text{   strat.\ numbers: }}\fwboxL{280pt}{\hspace{85pt}\{1,\!8,\!35,\!120,\!231,\!299,\!279,\!197,\!90,\!20\}}&\fwbox{50pt}{\chi=8}\\\hline\multicolumn{3}{|@{}l@{}|}{\fwbox{0pt}{\rule{0.pt}{70.75pt}}\raisebox{36.15pt}{$\fwboxL{355pt}{\hspace{5pt}\displaystyle\scalebox{0.85}{$\displaystyle\frac{\m{134}^2\m{456}}{\m{123}\m{124}\m{145}\m{146}\m{345}\m{346}\m{356}\m{14\inter{23}{56}}}$}}$}}\\\hline\end{array}\,\begin{array}{@{}c@{}}\hspace{-3.95pt}\fig{-42.9pt}{0.75}{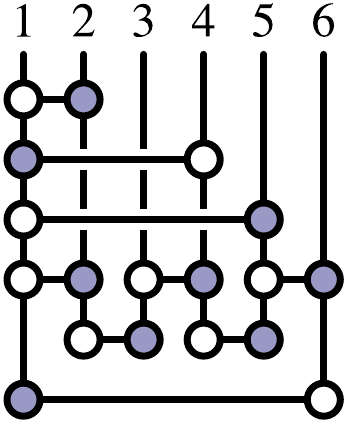}\end{array}\nonumber}}
\vspace{-2pt}\eq{\fwbox{430pt}{\begin{array}{|@{}l@{}|@{}l@{}|@{}r@{}|}\hline\fwboxL{25pt}{\,\,\Omega_{13}}&\fwboxL{0pt}{\text{   strat.\ numbers: }}\fwboxL{280pt}{\hspace{85pt}\{1,\!9,\!40,\!147,\!271,\!332,\!294,\!201,\!90,\!20\}}&\fwbox{50pt}{\chi=13}\\\hline\multicolumn{3}{|@{}l@{}|}{\fwbox{0pt}{\rule{0.pt}{55.75pt}}\raisebox{26.15pt}{$\fwboxL{355pt}{\hspace{5pt}\displaystyle\scalebox{0.85}{$\displaystyle\frac{\m{145}^2\m{234}^2}{\m{123}\m{124}\m{134}\m{146}\m{235}\m{245}\m{345}\m{456}\m{14\inter{23}{56}}}$}}$}}\\\hline\multicolumn{3}{@{}c@{}}{\rule{0pt}{10pt}}\\[-16pt]\end{array}\,\begin{array}{@{}c@{}}\\[-20pt]\hspace{-3.95pt}\fig{-42.9pt}{0.75}{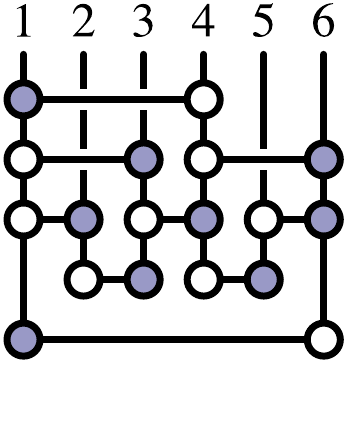}\\[-20pt]\end{array}\vspace{-50pt}\nonumber}}
\vspace{-2pt}\eq{\fwbox{430pt}{\begin{array}{|@{}l@{}|@{}l@{}|@{}r@{}|}\hline\fwboxL{25pt}{\,\,\Omega_{14}}&\fwboxL{0pt}{\text{   strat.\ numbers: }}\fwboxL{280pt}{\hspace{85pt}\{1,\!7,\!25,\!93,\!186,\!259,\!260,\!193,\!90,\!20\}}&\fwbox{50pt}{\chi=10}\\\hline\multicolumn{3}{|@{}l@{}|}{\fwbox{0pt}{\rule{0.pt}{70.75pt}}\raisebox{36.15pt}{$\fwboxL{355pt}{\hspace{5pt}\displaystyle\scalebox{0.85}{$\displaystyle\frac{\m{125}^3}{\m{123}\m{124}\m{135}\m{156}\m{245}\m{256}\m{4\inter{12}{35}\inter{16}{25}}}$}}$}}\\\hline\end{array}\,\begin{array}{@{}c@{}}\hspace{-3.95pt}\fig{-42.9pt}{0.75}{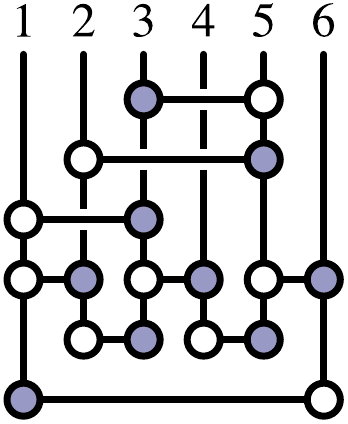}\end{array}\nonumber}}
\vspace{-2pt}\eq{\fwbox{430pt}{\begin{array}{|@{}l@{}|@{}l@{}|@{}r@{}|}\hline\fwboxL{25pt}{\,\,\Omega_{15}}&\fwboxL{0pt}{\text{   strat.\ numbers: }}\fwboxL{280pt}{\hspace{85pt}\{1,\!7,\!29,\!103,\!206,\!281,\!272,\!196,\!90,\!20\}}&\fwbox{50pt}{\chi=9}\\\hline\multicolumn{3}{|@{}l@{}|}{\fwbox{0pt}{\rule{0.pt}{70.75pt}}\raisebox{36.15pt}{$\fwboxL{355pt}{\hspace{5pt}\displaystyle\scalebox{0.85}{$\displaystyle\frac{\m{125}\m{235}^2}{\m{123}\m{135}\m{156}\m{234}\m{245}\m{256}\m{6\inter{15}{23}\inter{25}{34}}}$}}$}}\\\hline\end{array}\,\begin{array}{@{}c@{}}\hspace{-3.95pt}\fig{-42.9pt}{0.75}{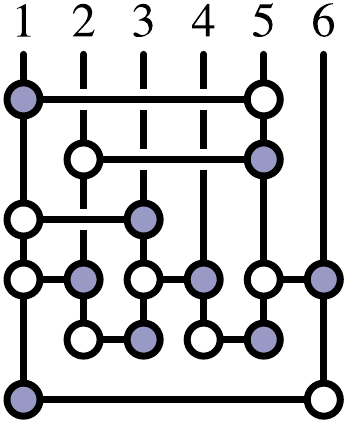}\end{array}\nonumber}}
\vspace{-2pt}\eq{\fwbox{430pt}{\begin{array}{|@{}l@{}|@{}l@{}|@{}r@{}|}\hline\fwboxL{25pt}{\,\,\Omega_{16}}&\fwboxL{0pt}{\text{   strat.\ numbers: }}\fwboxL{280pt}{\hspace{85pt}\{1,\!5,\!26,\!94,\!187,\!259,\!260,\!193,\!90,\!20\}}&\fwbox{50pt}{\chi=7}\\\hline\multicolumn{3}{|@{}l@{}|}{\fwbox{0pt}{\rule{0.pt}{70.75pt}}\raisebox{36.15pt}{$\fwboxL{355pt}{\hspace{5pt}\displaystyle\scalebox{0.85}{$\displaystyle\frac{\m{123}\m{146}}{\m{124}\m{165}\m{236}\m{14\inter{23}{56}}\m{6\inter{14}{23}\inter{15}{34}}}$}}$}}\\\hline\end{array}\,\begin{array}{@{}c@{}}\hspace{-3.95pt}\fig{-42.9pt}{0.75}{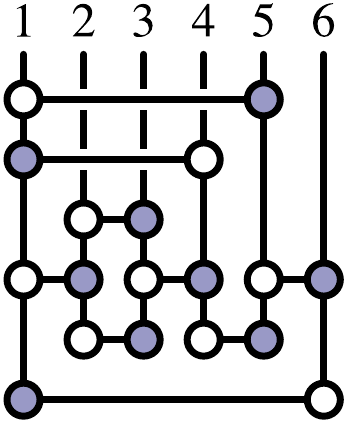}\end{array}\nonumber}}
\vspace{-2pt}\eq{\fwbox{430pt}{\begin{array}{|@{}l@{}|@{}l@{}|@{}r@{}|}\hline\fwboxL{25pt}{\,\,\Omega_{17}}&\fwboxL{0pt}{\text{   strat.\ numbers: }}\fwboxL{280pt}{\hspace{85pt}\{1,\!6,\!31,\!116,\!220,\!288,\!275,\!197,\!90,\!20\}}&\fwbox{50pt}{\chi=10}\\\hline\multicolumn{3}{|@{}l@{}|}{\fwbox{0pt}{\rule{0.pt}{70.75pt}}\raisebox{36.15pt}{$\fwboxL{355pt}{\hspace{5pt}\displaystyle\scalebox{0.85}{$\displaystyle\frac{\m{124}\m{234}\m{235}}{\m{123}\m{134}\m{245}\m{256}\m{14\inter{23}{56}}\m{6\inter{14}{23}\inter{25}{34}}}$}}$}}\\\hline\end{array}\,\begin{array}{@{}c@{}}\hspace{-3.95pt}\fig{-42.9pt}{0.75}{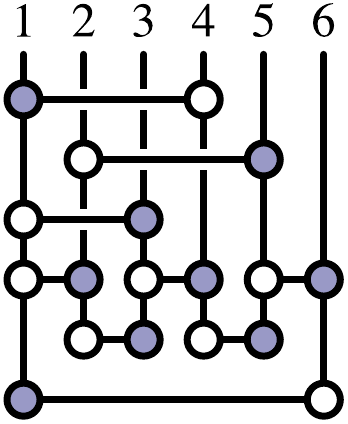}\end{array}\nonumber}}
\vspace{-2pt}\eq{\fwbox{430pt}{\begin{array}{|@{}l@{}|@{}l@{}|@{}r@{}|}\hline\fwboxL{25pt}{\,\,\Omega_{18}}&\fwboxL{0pt}{\text{   strat.\ numbers: }}\fwboxL{280pt}{\hspace{85pt}\{1,\!7,\!39,\!151,\!280,\!341,\!299,\!202,\!90,\!20\}}&\fwbox{50pt}{\chi=12}\\\hline\multicolumn{3}{|@{}l@{}|}{\fwbox{0pt}{\rule{0.pt}{70.75pt}}\raisebox{36.15pt}{$\fwboxL{355pt}{\hspace{5pt}\displaystyle\scalebox{0.85}{$\displaystyle\frac{\m{123}^2\m{145}\m{146}^2\m{234}}{\m{124}\m{134}\m{156}\m{236}\m{14\inter{23}{56}}\m{6\inter{14}{23}\inter{15}{24}}\m{6\inter{14}{23}\inter{15}{34}}}$}}$}}\\\hline\end{array}\,\begin{array}{@{}c@{}}\hspace{-3.95pt}\fig{-42.9pt}{0.75}{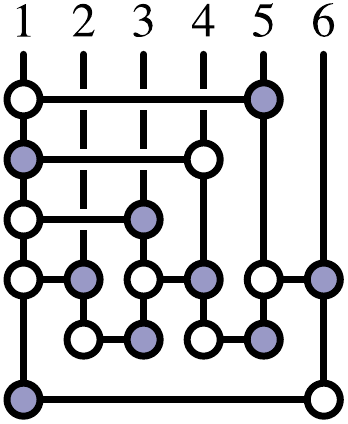}\end{array}\nonumber}}
\vspace{-2pt}\eq{\fwbox{430pt}{\begin{array}{|@{}l@{}|@{}l@{}|@{}r@{}|}\hline\fwboxL{25pt}{\,\,\Omega_{19}}&\fwboxL{0pt}{\text{   strat.\ numbers: }}\fwboxL{280pt}{\hspace{85pt}\{1,\!8,\!36,\!117,\!223,\!293,\!277,\!197,\!90,\!20\}}&\fwbox{50pt}{\chi=8}\\\hline\multicolumn{3}{|@{}l@{}|}{\fwbox{0pt}{\rule{0.pt}{70.75pt}}\raisebox{36.15pt}{$\fwboxL{355pt}{\hspace{5pt}\displaystyle\scalebox{0.85}{$\displaystyle\frac{\m{6\inter{12}{34}\inter{14}{23}}^2}{\m{123}\m{124}\m{125}\m{146}\m{236}\m{346}\m{6\inter{14}{23}\inter{15}{34}}\m{6\inter{14}{23}\inter{25}{34}}}$}}$}}\\\hline\end{array}\,\begin{array}{@{}c@{}}\hspace{-3.95pt}\fig{-42.9pt}{0.75}{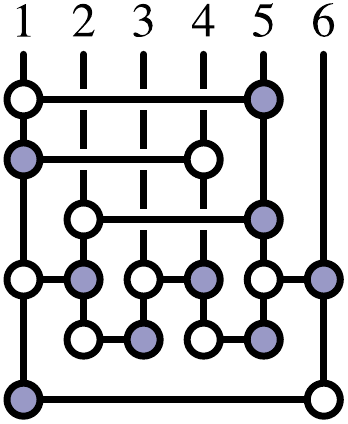}\end{array}\nonumber}}
\vspace{-2pt}\eq{\fwbox{430pt}{\begin{array}{|@{}l@{}|@{}l@{}|@{}r@{}|}\hline\fwboxL{25pt}{\,\,\Omega_{20}}&\fwboxL{0pt}{\text{   strat.\ numbers: }}\fwboxL{280pt}{\hspace{85pt}\{1,\!3,\!15,\!66,\!153,\!231,\!246,\!189,\!90,\!20\}}&\fwbox{50pt}{\chi=4}\\\hline\multicolumn{3}{|@{}l@{}|}{\fwbox{0pt}{\rule{0.pt}{70.75pt}}\raisebox{36.15pt}{$\fwboxL{355pt}{\hspace{5pt}\displaystyle\scalebox{0.85}{$\displaystyle\frac{\m{124}\m{236}}{2z\Delta\m{12\hat{3}}\m{146}\m{256}\m{56\inter{14}{2\hat{3}}}}$}}\fwboxR{0pt}{\scalebox{0.75}{$\begin{array}{@{}c@{$\,$}l@{}}\\\widehat{3}&\equiv 3\pl z\,6\\z&\equiv \m{234}/\m{246}-\Delta/a\\\Delta&\equiv\sqrt{\m{146}\m{236}\m{245}\m{256}\m{46\inter{34}{12}}}\\a&\equiv\fwboxL{225pt}{\m{146}\m{246}\m{256}}\end{array}\;\;$}}
$}}\\\hline\end{array}\,\begin{array}{@{}c@{}}\hspace{-3.95pt}\fig{-42.9pt}{0.75}{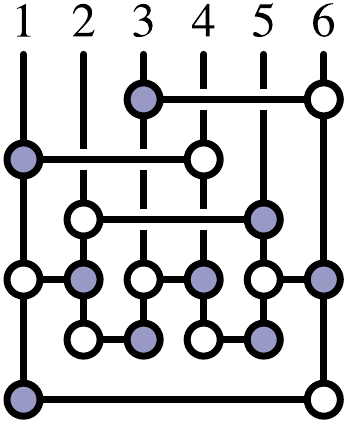}\end{array}\nonumber}}
\vspace{-2pt}\eq{\fwbox{430pt}{\begin{array}{|@{}l@{}|@{}l@{}|@{}r@{}|}\hline\fwboxL{25pt}{\,\,\Omega_{21}}&\fwboxL{0pt}{\text{   strat.\ numbers: }}\fwboxL{280pt}{\hspace{85pt}\{1,\!6,\!28,\!108,\!216,\!289,\!276,\!197,\!90,\!20\}}&\fwbox{50pt}{\chi=9}\\\hline\multicolumn{3}{|@{}l@{}|}{\fwbox{0pt}{\rule{0.pt}{70.75pt}}\raisebox{36.15pt}{$\fwboxL{355pt}{\hspace{5pt}\displaystyle\scalebox{0.85}{$\displaystyle\frac{\m{124}\m{2\hat{3}6}}{z\Delta\m{123}\m{146}\m{256}\m{56\inter{14}{2\hat{3}}}}$}}\fwboxR{0pt}{\scalebox{0.75}{$\begin{array}{@{}c@{$\,$}l@{}}\\[-10pt]\widehat{3}&\equiv 3\pl z\,1\\z&\equiv(b-\Delta)/a\\\Delta&\equiv\sqrt{b^2-4\,c}\\a&\equiv2\,\m{124}\m{125}\m{146}\\b&\equiv\m{6\inter{14}{23}\inter{14}{25}}\mi\m{124}\m{16\inter{25}{34}}\\c&\equiv\fwboxL{225pt}{\m{124}\m{125}\m{146}\m{6\inter{23}{14}\inter{25}{34}}}\end{array}\;\;$}}
$}}\\\hline\end{array}\,\begin{array}{@{}c@{}}\hspace{-3.95pt}\fig{-42.9pt}{0.75}{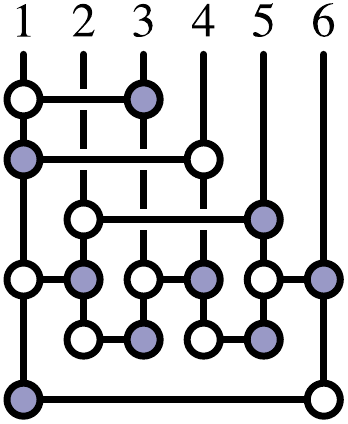}\end{array}\nonumber}}
\vspace{-2pt}\eq{\fwbox{430pt}{\begin{array}{|@{}l@{}|@{}l@{}|@{}r@{}|}\hline\fwboxL{25pt}{\,\,\Omega_{22}}&\fwboxL{0pt}{\text{   strat.\ numbers: }}\fwboxL{280pt}{\hspace{85pt}\{1,\!9,\!38,\!132,\!236,\!298,\!279,\!198,\!90,\!20\}}&\fwbox{50pt}{\chi=13}\\\hline\multicolumn{3}{|@{}l@{}|}{\fwbox{0pt}{\rule{0.pt}{70.75pt}}\raisebox{36.15pt}{$\fwboxL{355pt}{\hspace{5pt}\displaystyle\scalebox{0.85}{$\displaystyle\frac{\m{124}\m{\hat{2}36}}{z\Delta\m{123}\m{146}\m{\hat{2}56}\m{56\inter{14}{\hat{2}3}}}$}}\fwboxR{0pt}{\scalebox{0.75}{$\begin{array}{@{}c@{$\,$}l@{}}\\[-10pt]\widehat{2}&\equiv 2\pl z\,1\\z&\equiv(b+\Delta)/a\\\Delta&\equiv\sqrt{b^2-4\,c}\\a&\equiv2\,\m{134}^2\m{156}\\b&\equiv\m{6\inter{14}{23}\inter{15}{34}}\mi\m{134}\m{16\inter{25}{34}}\\c&\equiv\fwboxL{225pt}{\m{134}^2\m{156}\m{6\inter{23}{14}\inter{25}{34}}}\end{array}\;\;$}}
$}}\\\hline\end{array}\,\begin{array}{@{}c@{}}\hspace{-3.95pt}\fig{-42.9pt}{0.75}{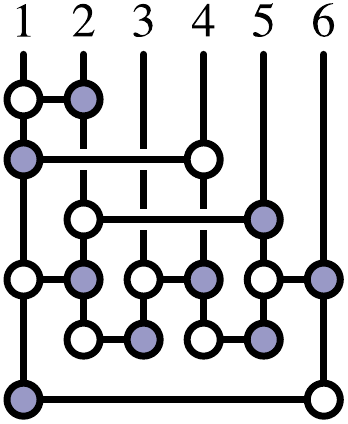}\end{array}\nonumber}}
\vspace{-2pt}\eq{\fwbox{430pt}{\begin{array}{|@{}l@{}|@{}l@{}|@{}r@{}|}\hline\fwboxL{25pt}{\,\,\Omega_{23}}&\fwboxL{0pt}{\text{   strat.\ nos: }}\fwboxL{254pt}{\hspace{60pt}\{1,\!12,\!54,\!166,\!348,\!420,\!339,\!210,\!90,\!20\}}&\fwbox{50pt}{\chi=-4}\\\hline\multicolumn{3}{|@{}l@{}|}{\fwbox{0pt}{\rule{0.pt}{70.75pt}}\raisebox{36.15pt}{$\fwboxL{329pt}{\hspace{5pt}\displaystyle\scalebox{0.85}{$\displaystyle\frac{\m{61\inter{25}{34}}^3}{\m{125}\m{126}\m{134}\m{136}\m{146}\m{156}\m{234}\m{235}\m{245}\m{256}\m{345}\m{346}}$}}$}}\\\hline\end{array}\,\begin{array}{@{}c@{}}\hspace{-3.95pt}\fig{-42.9pt}{0.75}{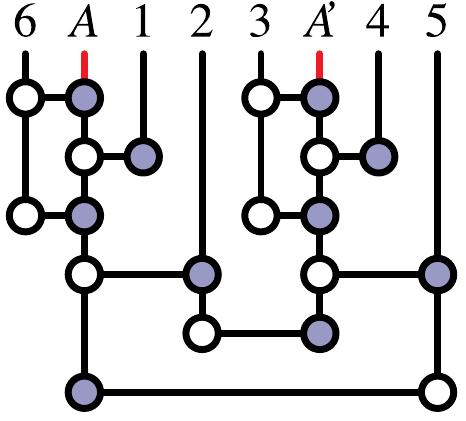}\end{array}}\nonumber}
\vspace{-2pt}\eq{\fwbox{430pt}{\begin{array}{|@{}l@{}|@{}l@{}|@{}r@{}|}\hline\fwboxL{25pt}{\,\,\Omega_{24}}&\fwboxL{0pt}{\text{   strat.\ nos: }}\fwboxL{254pt}{\hspace{60pt}\{1,\!6,\!21,\!74,\!157,\!232,\!246,\!189,\!90,\!20\}}&\fwbox{50pt}{\chi=6}\\\hline\multicolumn{3}{|@{}l@{}|}{\fwbox{0pt}{\rule{0.pt}{70.75pt}}\raisebox{36.15pt}{$\fwboxL{329pt}{\hspace{5pt}\displaystyle\scalebox{0.85}{$\displaystyle\frac{\m{456}^2}{\Delta\m{145}\m{164}\m{245}\m{256}\m{364}\m{356}}$}}\fwboxR{0pt}{\scalebox{0.75}{$\begin{array}{@{}c@{$\,$}l@{}}\Delta&\equiv\sqrt{b^2-4\,c}\\b&\equiv\m{124}\m{356}\pl\m{146}\m{235}\mi\m{156}\m{234}\\c&\equiv\fwboxL{191pt}{\m{126}\m{145}\m{234}\m{356}}\end{array}\;\;$}}$}}\\\hline\end{array}\,\begin{array}{@{}c@{}}\hspace{-3.95pt}\fig{-42.9pt}{0.75}{top_form_diagram_24c}\end{array}}\nonumber}

\newpage
\section{Leading Singularities and Residue Theorems in $G(3,6)$}\label{appendix:leading_singularities}\vspace{-6pt}
In this appendix, we provide representatives of all inequivalent 8-dimensional on-shell varieties of $G(3,6)$. There are only 10 of these---generated as co-dimension one boundaries of the top-forms given above. 

These on-shell varieties are of particular interest in physics, corresponding to leading singularities of $6$-particle `NMHV' loop amplitudes. On-shell functions are represented by on-shell varieties as follows:
\eq{f(\lambda,\widetilde\lambda,\widetilde\eta)=\int\!\Omega_C\,\,\delta^{3\times4}\hspace{-0.5pt}\big(C\!\cdot\!\widetilde{\eta}\big)\delta^{3\times2}\hspace{-0.5pt}\big(C\!\cdot\!\widetilde\lambda\big)\delta^{2\times3}\hspace{-0.5pt}\big(\lambda\!\cdot\!C^{\perp}\big).\label{representation_of_on_shell_functions}}
Here, the volume form $\Omega$ is that of some cluster variety of $G(3,6)$; when integrated against the kinematical $\delta$-functions, these forms become ordinary (rational) functions of $\lambda,\widetilde\lambda,\widetilde\eta$, the kinematical data describing the external particles. We will not review the details of this story here, but only note that the 12 (bosonic) $\delta$-functions in (\ref{representation_of_on_shell_functions}) impose 8 independent constraints on $C$ together with the 4 constraints of momentum conservation, $\delta^{2\times2}\hspace{-0.5pt}\big(\lambda\!\cdot\!\widetilde\lambda\big)$. Thus, the volume-form of any 8-dimensional variety is converted via (\ref{representation_of_on_shell_functions}) to a rational function of the kinematical data. 

When there is a unique solution to the $\delta$-functions constraints, $C\!\mapsto\!C^*$, the on-shell function is directly a rational function of $C^*(\lambda,\widetilde\lambda)$. This is the case for 7 of the 10 inequivalent functions, allowing us to write explicit formulae for each in terms of $(\lambda,\widetilde\lambda)$. For the remaining 3, we provide a concrete coordinate chart for the variety, in terms of which the function is represented according to (\ref{representation_of_on_shell_functions}).

\eq{\fwbox{430pt}{\begin{array}{|@{}l@{}|@{}l@{}|@{}r@{}|}\hline\multicolumn{3}{|@{}l@{}|}{\fwbox{0pt}{\rule{0.pt}{36.75pt}}\raisebox{16.15pt}{$\fwboxL{355pt}{\hspace{5pt}\displaystyle\scalebox{0.85}{$\displaystyle f_{1}\!\equiv\!\!\!\!\oint\limits_{\m{123}=0}\!\!\!\!\!\Omega_{1}\!=\!\left.\frac{\!\,\delta^{3\times4}\hspace{-0.5pt}\big(C^*\!\!\cdot\!\widetilde\eta\big)\delta^{2\times2}\hspace{-0.5pt}\big(\lambda\!\cdot\!\widetilde\lambda\big)}{\m{234}\m{345}\m{456}\m{561}\m{612}}\right|_{C^*}$}}\fwboxR{0pt}{\scalebox{0.7}{$\displaystyle C^*\!\equiv\!\left(\!\!\begin{array}{@{}c@{}c@{}c@{}c@{}c@{}c@{}}\lambda_1^1&\lambda_2^1&\lambda_3^1&\lambda_4^1&\lambda_5^1&\lambda_6^1\\\lambda_1^2&\lambda_2^2&\lambda_3^2&\lambda_4^2&\lambda_5^2&\lambda_6^2\\[-2pt]\fwbox{20pt}{{\color{dim}0}}&\fwbox{20pt}{{\color{dim}0}}&\fwbox{20pt}{{\color{dim}0}}&\fwbox{20pt}{\sb{56}}&\fwbox{20pt}{\sb{64}}&\fwbox{20pt}{\sb{45}}\end{array}\!\right)$}\hspace{0pt}}$}}\\[-0pt]\multicolumn{3}{|@{}l@{}|}{\fwbox{0pt}{\rule[-10pt]{0.0pt}{20.75pt}}\raisebox{-0.15pt}[15pt][14pt]{$\fwboxL{355pt}{\hspace{5pt}\displaystyle\scalebox{0.85}{$\displaystyle \phantom{f_{1}\!\equiv\!\!\!\!\oint\limits_{\m{123}=0}\!\!\!\!\!\Omega_{1}\!}=\!\frac{\!\,\delta^{3\times4}\hspace{-0.5pt}\big(C^*\!\!\cdot\!\widetilde\eta\big)\delta^{2\times2}\hspace{-0.5pt}\big(\lambda\!\cdot\!\widetilde\lambda\big)}{\ab{23}\sb{56}\asb{3}{4}{5}{6}s_{456}\asb{1}{5}{6}{4}\ab{12}\sb{45}}$}}$}}\\\hline\multicolumn{3}{@{}c@{}}{~}\\[-15pt]\end{array}\,\begin{array}{@{}c@{}}\\[-22pt]\hspace{-3.95pt}\fig{-42.9pt}{0.75}{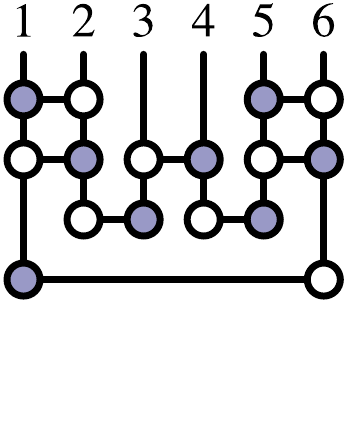}\\[-32pt]\end{array}\nonumber}}
\eq{\fwbox{430pt}{\begin{array}{|@{}l@{}|@{}l@{}|@{}r@{}|}\hline\multicolumn{3}{|@{}l@{}|}{\fwbox{0pt}{\rule{0.pt}{42.75pt}}\raisebox{16.15pt}{$\fwboxL{355pt}{\hspace{5pt}\displaystyle\scalebox{0.85}{$\displaystyle f_{2}\!\equiv\!\!\!\!\oint\limits_{\m{123}=0}\!\!\!\!\!\Omega_{2}\!=\!\left.\frac{\m{235}\,\delta^{3\times4}\hspace{-0.5pt}\big(C^*\!\!\cdot\!\widetilde\eta\big)\delta^{2\times2}\hspace{-0.5pt}\big(\lambda\!\cdot\!\widetilde\lambda\big)}{\m{136}\m{156}\m{234}\m{245}\m{256}\m{345}}\right|_{C^*}$}}\fwboxR{0pt}{\scalebox{0.7}{$\displaystyle C^*\!\equiv\!\left(\!\!\begin{array}{@{}c@{}c@{}c@{}c@{}c@{}c@{}}\lambda_1^1&\lambda_2^1&\lambda_3^1&\lambda_4^1&\lambda_5^1&\lambda_6^1\\\lambda_1^2&\lambda_2^2&\lambda_3^2&\lambda_4^2&\lambda_5^2&\lambda_6^2\\[-2pt]\fwbox{20pt}{{\color{dim}0}}&\fwbox{20pt}{{\color{dim}0}}&\fwbox{20pt}{{\color{dim}0}}&\fwbox{20pt}{\sb{56}}&\fwbox{20pt}{\sb{64}}&\fwbox{20pt}{\sb{45}}\end{array}\!\right)$}\hspace{0pt}}$}}\\[-5pt]\multicolumn{3}{|@{}l@{}|}{\fwbox{0pt}{\rule[-10pt]{0.0pt}{20.75pt}}\raisebox{-0.15pt}[18pt][17pt]{$\fwboxL{355pt}{\hspace{5pt}\displaystyle\scalebox{0.85}{$\displaystyle \phantom{f_{2}\!\equiv\!\!\!\!\oint\limits_{\m{123}=0}\!\!\!\!\!\Omega_{2}\!}\hspace{-0pt}=\!\frac{\ab{23}\sb{64}\,\delta^{3\times4}\hspace{-0.5pt}\big(C^*\!\!\cdot\!\widetilde\eta\big)\delta^{2\times2}\hspace{-0.5pt}\big(\lambda\!\cdot\!\widetilde\lambda\big)}{\ab{13}\sb{45}\asb{1}{5}{6}{4}\ab{23}\sb{56}\asb{2}{4}{5}{6}\asb{2}{5}{6}{4}\asb{3}{4}{5}{6}}$}}$}}\\\hline\multicolumn{3}{@{}c@{}}{~}\\[-15pt]\end{array}\,\begin{array}{@{}c@{}}\\[-44pt]\hspace{-3.95pt}\fig{-42.9pt}{0.75}{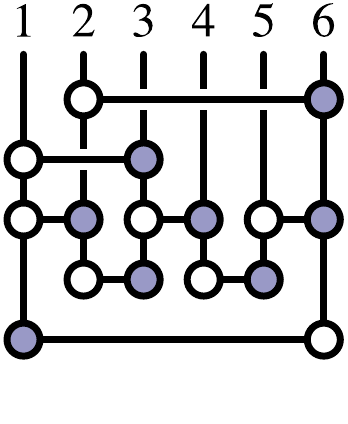}\\[-42pt]\end{array}\nonumber}}
\eq{\fwbox{430pt}{\begin{array}{|@{}l@{}|@{}l@{}|@{}r@{}|}\hline\multicolumn{3}{|@{}l@{}|}{\fwbox{0pt}{\rule{0.pt}{36.75pt}}\raisebox{16.15pt}{$\fwboxL{355pt}{\hspace{5pt}\displaystyle\scalebox{0.85}{$\displaystyle f_{3}\!\equiv\!\!\!\!\oint\limits_{\m{123}=0}\!\!\!\!\!\Omega_{4}\!=\!\left.\frac{\m{145}\,\delta^{3\times4}\hspace{-0.5pt}\big(C^*\!\!\cdot\!\widetilde\eta\big)\delta^{2\times2}\hspace{-0.5pt}\big(\lambda\!\cdot\!\widetilde\lambda\big)}{\m{124}\m{136}\m{156}\m{245}\m{345}\m{456}}\right|_{C^*}$}}\fwboxR{0pt}{\scalebox{0.7}{$\displaystyle C^*\!\equiv\!\left(\!\!\begin{array}{@{}c@{}c@{}c@{}c@{}c@{}c@{}}\lambda_1^1&\lambda_2^1&\lambda_3^1&\lambda_4^1&\lambda_5^1&\lambda_6^1\\\lambda_1^2&\lambda_2^2&\lambda_3^2&\lambda_4^2&\lambda_5^2&\lambda_6^2\\[-2pt]\fwbox{20pt}{{\color{dim}0}}&\fwbox{20pt}{{\color{dim}0}}&\fwbox{20pt}{{\color{dim}0}}&\fwbox{20pt}{\sb{56}}&\fwbox{20pt}{\sb{64}}&\fwbox{20pt}{\sb{45}}\end{array}\!\right)$}\hspace{0pt}}$}}\\[-0pt]\multicolumn{3}{|@{}l@{}|}{\fwbox{0pt}{\rule[-10pt]{0.0pt}{20.75pt}}\raisebox{-0.15pt}[15pt][14pt]{$\fwboxL{355pt}{\hspace{5pt}\displaystyle\scalebox{0.85}{$\displaystyle \phantom{f_{3}\!\equiv\!\!\!\!\oint\limits_{\m{123}=0}\!\!\!\!\!\Omega_{4}\!}=\!\frac{\asb{1}{4}{5}{6}\,\delta^{3\times4}\hspace{-0.5pt}\big(C^*\!\!\cdot\!\widetilde\eta\big)\delta^{2\times2}\hspace{-0.5pt}\big(\lambda\!\cdot\!\widetilde\lambda\big)}{\ab{12}\sb{56}\ab{13}\sb{45}\asb{1}{5}{6}{4}\asb{2}{4}{5}{6}\asb{3}{4}{5}{6}s_{456}}$}}$}}\\\hline\multicolumn{3}{@{}c@{}}{~}\\[-15pt]\end{array}\,\begin{array}{@{}c@{}}\\[-22pt]\hspace{-3.95pt}\fig{-42.9pt}{0.75}{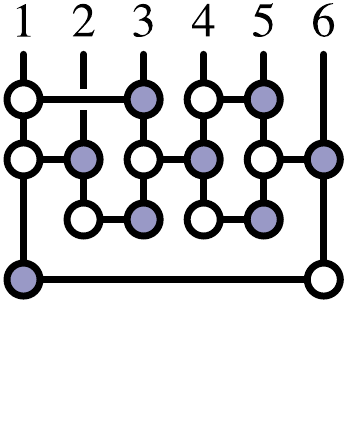}\\[-32pt]\end{array}\nonumber}}
\eq{\fwbox{430pt}{\begin{array}{|@{}l@{}|@{}l@{}|@{}r@{}|}\hline\multicolumn{3}{|@{}l@{}|}{\fwbox{0pt}{\rule{0.pt}{42.75pt}}\raisebox{16.15pt}{$\fwboxL{355pt}{\hspace{5pt}\displaystyle\scalebox{0.85}{$\displaystyle f_{4}\!\equiv\!\!\!\!\oint\limits_{\m{123}=0}\!\!\!\!\!\Omega_{5}\!=\!\left.\frac{\m{135}\,\delta^{3\times4}\hspace{-0.5pt}\big(C^*\!\!\cdot\!\widetilde\eta\big)\delta^{2\times2}\hspace{-0.5pt}\big(\lambda\!\cdot\!\widetilde\lambda\big)}{\m{124}\m{145}\m{156}\m{236}\m{345}\m{356}}\right|_{C^*}$}}\fwboxR{0pt}{\scalebox{0.7}{$\displaystyle C^*\!\equiv\!\left(\!\!\begin{array}{@{}c@{}c@{}c@{}c@{}c@{}c@{}}\lambda_1^1&\lambda_2^1&\lambda_3^1&\lambda_4^1&\lambda_5^1&\lambda_6^1\\\lambda_1^2&\lambda_2^2&\lambda_3^2&\lambda_4^2&\lambda_5^2&\lambda_6^2\\[-2pt]\fwbox{20pt}{{\color{dim}0}}&\fwbox{20pt}{{\color{dim}0}}&\fwbox{20pt}{{\color{dim}0}}&\fwbox{20pt}{\sb{56}}&\fwbox{20pt}{\sb{64}}&\fwbox{20pt}{\sb{45}}\end{array}\!\right)$}\hspace{0pt}}$}}\\[-5pt]\multicolumn{3}{|@{}l@{}|}{\fwbox{0pt}{\rule[-10pt]{0.0pt}{20.75pt}}\raisebox{-0.15pt}[18pt][17pt]{$\fwboxL{355pt}{\hspace{5pt}\displaystyle\scalebox{0.85}{$\displaystyle \phantom{f_{4}\!\equiv\!\!\!\!\oint\limits_{\m{123}=0}\!\!\!\!\!\Omega_{5}\!}\hspace{-0pt}=\!\frac{\ab{13}\sb{64}\,\delta^{3\times4}\hspace{-0.5pt}\big(C^*\!\!\cdot\!\widetilde\eta\big)\delta^{2\times2}\hspace{-0.5pt}\big(\lambda\!\cdot\!\widetilde\lambda\big)}{\ab{12}\sb{56}\asb{1}{4}{5}{6}\asb{1}{5}{6}{4}\ab{23}\sb{45}\asb{3}{4}{5}{6}\asb{3}{5}{6}{4}}$}}$}}\\\hline\multicolumn{3}{@{}c@{}}{~}\\[-15pt]\end{array}\,\begin{array}{@{}c@{}}\\[-44pt]\hspace{-3.95pt}\fig{-42.9pt}{0.75}{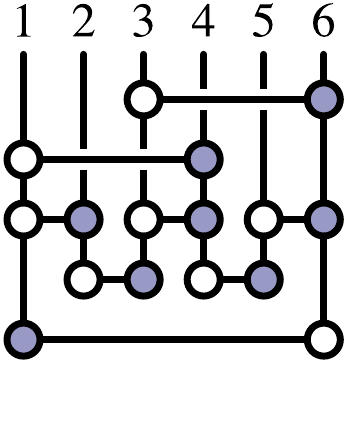}\\[-42pt]\end{array}\nonumber}}
\eq{\fwbox{430pt}{\begin{array}{|@{}l@{}|@{}l@{}|@{}r@{}|}\hline\multicolumn{3}{|@{}l@{}|}{\fwbox{0pt}{\rule{0.pt}{42.75pt}}\raisebox{16.15pt}{$\fwboxL{355pt}{\hspace{5pt}\displaystyle\scalebox{0.85}{$\displaystyle f_{5}\!\equiv\!\!\!\!\oint\limits_{\m{123}=0}\!\!\!\!\!\Omega_{9}\!=\!\left.\frac{\m{125}\,\delta^{3\times4}\hspace{-0.5pt}\big(C^*\!\!\cdot\!\widetilde\eta\big)\delta^{2\times2}\hspace{-0.5pt}\big(\lambda\!\cdot\!\widetilde\lambda\big)}{\m{134}\m{156}\m{245}\m{256}\m{16\inter{25}{34}}}\right|_{C^*}$}}\fwboxR{0pt}{\scalebox{0.7}{$\displaystyle C^*\!\equiv\!\left(\!\!\begin{array}{@{}c@{}c@{}c@{}c@{}c@{}c@{}}\lambda_1^1&\lambda_2^1&\lambda_3^1&\lambda_4^1&\lambda_5^1&\lambda_6^1\\\lambda_1^2&\lambda_2^2&\lambda_3^2&\lambda_4^2&\lambda_5^2&\lambda_6^2\\[-2pt]\fwbox{20pt}{{\color{dim}0}}&\fwbox{20pt}{{\color{dim}0}}&\fwbox{20pt}{{\color{dim}0}}&\fwbox{20pt}{\sb{56}}&\fwbox{20pt}{\sb{64}}&\fwbox{20pt}{\sb{45}}\end{array}\!\right)$}\hspace{0pt}}$}}\\[-5pt]\multicolumn{3}{|@{}l@{}|}{\fwbox{0pt}{\rule[-10pt]{0.0pt}{20.75pt}}\raisebox{-0.15pt}[18pt][17pt]{$\fwboxL{355pt}{\hspace{5pt}\displaystyle\scalebox{0.85}{$\displaystyle \phantom{f_{5}\!\equiv\!\!\!\!\oint\limits_{\m{123}=0}\!\!\!\!\!\Omega_{9}\!}\hspace{-20pt}=\!\frac{\ab{12}\sb{64}\,\delta^{3\times4}\hspace{-0.5pt}\big(C^*\!\!\cdot\!\widetilde\eta\big)\delta^{2\times2}\hspace{-0.5pt}\big(\lambda\!\cdot\!\widetilde\lambda\big)}{\ab{13}\sb{56}\asb{1}{5}{6}{4}\asb{2}{4}{5}{6}\asb{2}{5}{6}{4}\big(\hspace{-0.5pt}\ab{23}\sb{56}\asb{1}{5}{6}{4}\mi\ab{12}\sb{45}\asb{3}{4}{5}{6}\hspace{-0.5pt}\big)}$}}$}}\\\hline\multicolumn{3}{@{}c@{}}{~}\\[-15pt]\end{array}\,\begin{array}{@{}c@{}}\\[-44pt]\hspace{-3.95pt}\fig{-42.9pt}{0.75}{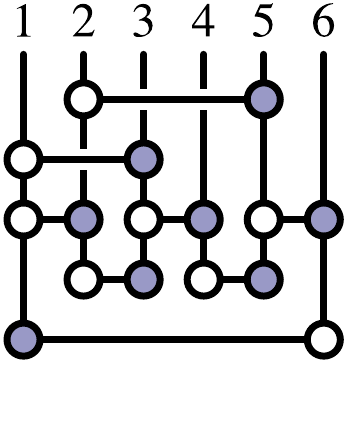}\\[-42pt]\end{array}\nonumber}}
\eq{\fwbox{430pt}{\begin{array}{|@{}l@{}|@{}l@{}|@{}r@{}|}\hline\multicolumn{3}{|@{}l@{}|}{\fwbox{0pt}{\rule{0.pt}{30.75pt}}\raisebox{16.15pt}{$\fwboxL{355pt}{\hspace{5pt}\displaystyle\scalebox{0.85}{$$}}\fwboxR{0pt}{\scalebox{0.7}{$\displaystyle C^*\!\equiv\!\left(\!\!\begin{array}{@{}c@{}c@{}c@{}c@{}c@{}c@{}}\lambda_1^1&\lambda_2^1&\lambda_3^1&\lambda_4^1&\lambda_5^1&\lambda_6^1\\\lambda_1^2&\lambda_2^2&\lambda_3^2&\lambda_4^2&\lambda_5^2&\lambda_6^2\\[-2pt]\fwbox{20pt}{{\color{dim}0}}&\fwbox{20pt}{{\color{dim}0}}&\fwbox{20pt}{{\color{dim}0}}&\fwbox{20pt}{\sb{56}}&\fwbox{20pt}{\sb{64}}&\fwbox{20pt}{\sb{45}}\end{array}\!\right)$}\hspace{0pt}}$}}\\[-15pt]\multicolumn{3}{|@{}l@{}|}{\fwbox{0pt}{\rule{0.pt}{30.75pt}}\raisebox{16.15pt}{$\fwboxL{355pt}{\hspace{5pt}\displaystyle\scalebox{0.85}{$\displaystyle f_{6}\!\equiv\!\!\!\!\oint\limits_{\m{123}=0}\!\!\!\!\!\Omega_{12}\!=\!\left.\frac{\m{134}^2\m{456}\,\delta^{3\times4}\hspace{-0.5pt}\big(C^*\!\!\cdot\!\widetilde\eta\big)\delta^{2\times2}\hspace{-0.5pt}\big(\lambda\!\cdot\!\widetilde\lambda\big)}{\m{124}\m{145}\m{146}\m{156}\m{234}\m{345}\m{346}\m{356}}\right|_{C^*}$}}$}}\\\multicolumn{3}{|@{}l@{}|}{\fwbox{0pt}{\rule[-10pt]{0.0pt}{20.75pt}}\raisebox{-0.15pt}[15pt][14pt]{$\fwboxL{355pt}{\hspace{5pt}\displaystyle\scalebox{0.85}{$\displaystyle \phantom{f_{6}\!\equiv\!\!\!\!\oint\limits_{\m{123}=0}\!\!\!\!\!\Omega_{12}\!}=\!\frac{\ab{13}^2 s_{456}\,\delta^{3\times4}\hspace{-0.5pt}\big(C^*\!\!\cdot\!\widetilde\eta\big)\delta^{2\times2}\hspace{-0.5pt}\big(\lambda\!\cdot\!\widetilde\lambda\big)}{\ab{12}\asb{1}{4}{5}{6}\asb{1}{4}{6}{5}\asb{1}{5}{6}{4}\ab{23}\asb{3}{4}{5}{6}\asb{3}{4}{6}{5}\asb{3}{5}{6}{4}}$}}$}}\\\hline\multicolumn{3}{@{}c@{}}{~}\\[-15pt]\end{array}\,\begin{array}{@{}c@{}}\\[-49pt]\hspace{-3.95pt}\fig{-42.9pt}{0.75}{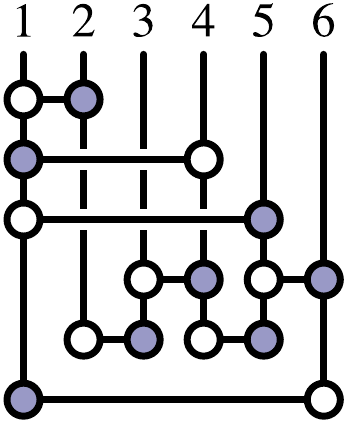}\\[-32pt]\end{array}\nonumber}}
\eq{\fwbox{430pt}{\begin{array}{|@{}l@{}|@{}l@{}|@{}r@{}|}\hline\multicolumn{3}{|@{}l@{}|}{\fwbox{0pt}{\rule{0.pt}{36.75pt}}\raisebox{16.15pt}{$\fwboxL{355pt}{\hspace{5pt}\displaystyle\scalebox{0.85}{$\displaystyle f_{7}\!\equiv\!\!\!\!\oint\limits_{\m{123}=0}\!\!\!\!\!\Omega_{13}\!=\!\left.\frac{\m{145}^2\,\delta^{3\times4}\hspace{-0.5pt}\big(C^*\!\!\cdot\!\widetilde\eta\big)\delta^{2\times2}\hspace{-0.5pt}\big(\lambda\!\cdot\!\widetilde\lambda\big)}{\m{125}\m{134}\m{146}\m{156}\m{245}\m{345}\m{456}}\right|_{C^*}$}}\fwboxR{0pt}{\scalebox{0.7}{$\displaystyle C^*\!\equiv\!\left(\!\!\begin{array}{@{}c@{}c@{}c@{}c@{}c@{}c@{}}\lambda_1^1&\lambda_2^1&\lambda_3^1&\lambda_4^1&\lambda_5^1&\lambda_6^1\\\lambda_1^2&\lambda_2^2&\lambda_3^2&\lambda_4^2&\lambda_5^2&\lambda_6^2\\[-2pt]\fwbox{20pt}{{\color{dim}0}}&\fwbox{20pt}{{\color{dim}0}}&\fwbox{20pt}{{\color{dim}0}}&\fwbox{20pt}{\sb{56}}&\fwbox{20pt}{\sb{64}}&\fwbox{20pt}{\sb{45}}\end{array}\!\right)$}\hspace{0pt}}$}}\\[-0pt]\multicolumn{3}{|@{}l@{}|}{\fwbox{0pt}{\rule[-10pt]{0.0pt}{20.75pt}}\raisebox{-0.15pt}[15pt][14pt]{$\fwboxL{355pt}{\hspace{5pt}\displaystyle\scalebox{0.85}{$\displaystyle \phantom{f_{7}\!\equiv\!\!\!\!\oint\limits_{\m{123}=0}\!\!\!\!\!\Omega_{13}\!}=\!\frac{\asb{1}{4}{5}{6}^2\,\delta^{3\times4}\hspace{-0.5pt}\big(C^*\!\!\cdot\!\widetilde\eta\big)\delta^{2\times2}\hspace{-0.5pt}\big(\lambda\!\cdot\!\widetilde\lambda\big)}{\ab{12}\sb{64}\ab{13}\sb{56}\asb{1}{4}{6}{5}\asb{1}{5}{6}{4}\asb{2}{4}{5}{6}\asb{3}{4}{5}{6}s_{456}}$}}$}}\\\hline\multicolumn{3}{@{}c@{}}{~}\\[-15pt]\end{array}\,\begin{array}{@{}c@{}}\\[-22pt]\hspace{-3.95pt}\fig{-42.9pt}{0.75}{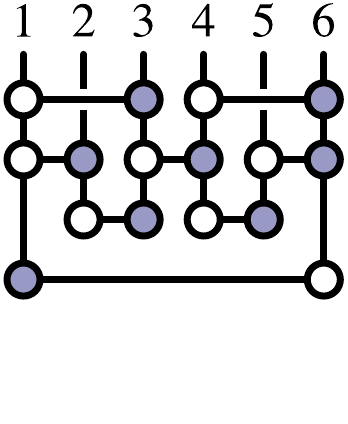}\\[-32pt]\end{array}\nonumber}}
\eq{\fwbox{430pt}{\begin{array}{|@{}l@{}|@{}l@{}|@{}r@{}|}\hline\multicolumn{3}{|@{}l@{}|}{\fwbox{0pt}{\rule{0.pt}{35.75pt}}\raisebox{16.15pt}{$\fwboxL{355pt}{\hspace{5pt}\displaystyle\scalebox{0.85}{$\displaystyle f_{8}\!\equiv\!\!\!\!\oint\limits_{\hspace{-12.5pt}\m{14(23)\scalebox{0.7}{$\newcap$}(56)}=0\hspace{-20.5pt}}\!\!\!\!\!\Omega_{16}\!=\int\!\frac{d\alpha_1}{\alpha_1}\!\wedge\!\cdots\!\wedge\!\frac{d\alpha_8}{\alpha_8}\,\,\delta^{3\times4}\hspace{-0.5pt}\big(C(\alpha)\!\cdot\!\widetilde{\eta}\big)\delta^{3\times2}\hspace{-0.5pt}\big(C(\alpha)\!\cdot\!\widetilde\lambda\big)\delta^{2\times3}\hspace{-0.5pt}\big(\lambda\!\cdot\!C^{\perp}(\alpha)\big)$}}$}}\\[-5pt]\multicolumn{3}{|@{}l@{}|}{\fwbox{0pt}{\rule[-10pt]{0.0pt}{20.75pt}}\raisebox{-0.15pt}[22pt][21pt]{$\fwbox{355pt}{\hspace{5pt}\displaystyle\scalebox{0.85}{$\displaystyle C(\alpha)\!\equiv\!\left(\!\begin{array}{@{$\,$}c@{$\;\,\,$}c@{$\;\,\,$}c@{$\;\,\,$}c@{$\;\,\,$}c@{$\;\,\,$}c@{$\,\,\,$}}1&\alpha_6&\alpha_6\,\alpha_7&{\color{dim}0}&{\color{dim}0}&\alpha_1\\{\color{dim}0}&1&\alpha_5\pl\alpha_7&{\color{dim}0}&\alpha_2&\alpha_2\,\alpha_4\\\alpha_8&{\color{dim}0}&{\color{dim}0}&1&\alpha_3&\alpha_3\,\alpha_4\end{array}\!\right)$}}$}}\\\hline\multicolumn{3}{@{}c@{}}{~}\\[-15pt]\end{array}\,\begin{array}{@{}c@{}}\\[-44pt]\hspace{-3.95pt}\fig{-42.9pt}{0.75}{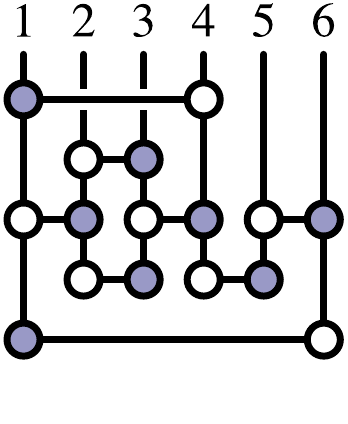}\\[-42pt]\end{array}\nonumber}}
\eq{\fwbox{430pt}{\begin{array}{|@{}l@{}|@{}l@{}|@{}r@{}|}\hline\multicolumn{3}{|@{}l@{}|}{\fwbox{0pt}{\rule{0.pt}{35.75pt}}\raisebox{16.15pt}{$\fwboxL{355pt}{\hspace{5pt}\displaystyle\scalebox{0.85}{$\displaystyle f_{9}\!\equiv\!\!\!\!\oint\limits_{\hspace{-12.5pt}\m{14(23)\scalebox{0.7}{$\newcap$}(56)}=0\hspace{-20.5pt}}\!\!\!\!\!\Omega_{18}\!=\int\!\frac{d\alpha_1}{\alpha_1}\!\wedge\!\cdots\!\wedge\!\frac{d\alpha_8}{\alpha_8}\,\,\delta^{3\times4}\hspace{-0.5pt}\big(C(\alpha)\!\cdot\!\widetilde{\eta}\big)\delta^{3\times2}\hspace{-0.5pt}\big(C(\alpha)\!\cdot\!\widetilde\lambda\big)\delta^{2\times3}\hspace{-0.5pt}\big(\lambda\!\cdot\!C^{\perp}(\alpha)\big)$}}$}}\\[-5pt]\multicolumn{3}{|@{}l@{}|}{\fwbox{0pt}{\rule[-10pt]{0.0pt}{20.75pt}}\raisebox{-0.15pt}[22pt][21pt]{$\fwbox{355pt}{\hspace{5pt}\displaystyle\scalebox{0.85}{$\displaystyle C(\alpha)\!\equiv\!\left(\!\begin{array}{@{$\,$}c@{$\;\,\,$}c@{$\;\,\,$}c@{$\;\,\,$}c@{$\;\,\,$}c@{$\;\,\,$}c@{$\,\,\,$}}1&\alpha_5&\alpha_7&{\color{dim}0}&{\color{dim}0}&\alpha_1\\{\color{dim}0}&1&\alpha_4&{\color{dim}0}&\alpha_2&\alpha_2\,\alpha_6\\\alpha_8&{\color{dim}0}&{\color{dim}0}&1&\alpha_3&\alpha_3\,\alpha_6\end{array}\!\right)$}}$}}\\\hline\multicolumn{3}{@{}c@{}}{~}\\[-15pt]\end{array}\,\begin{array}{@{}c@{}}\\[-44pt]\hspace{-3.95pt}\fig{-42.9pt}{0.75}{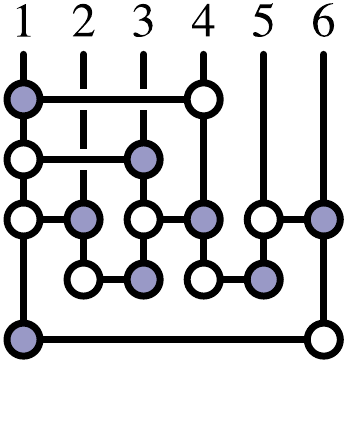}\\[-42pt]\end{array}\nonumber}}
\eq{\fwbox{430pt}{\begin{array}{|@{}l@{}|@{}l@{}|@{}r@{}|}\hline\multicolumn{3}{|@{}l@{}|}{\fwbox{0pt}{\rule{0.pt}{35.75pt}}\raisebox{16.15pt}{$\fwboxL{355pt}{\hspace{5pt}\displaystyle\scalebox{0.85}{$\displaystyle f_{10}\!\equiv\!\!\!\!\oint\limits_{\phantom{\hspace{-12.5pt}\m{14(23)\scalebox{0.7}{$\newcap$}(56)}=0\hspace{-20.5pt}}\fwbox{0pt}{\hspace{-30pt}z=0}}\!\!\!\!\!\Omega_{20}\!=\int\!\frac{d\alpha_1}{\alpha_1}\!\wedge\!\cdots\!\wedge\!\frac{d\alpha_8}{\alpha_8}\,\,\delta^{3\times4}\hspace{-0.5pt}\big(C(\alpha)\!\cdot\!\widetilde{\eta}\big)\delta^{3\times2}\hspace{-0.5pt}\big(C(\alpha)\!\cdot\!\widetilde\lambda\big)\delta^{2\times3}\hspace{-0.5pt}\big(\lambda\!\cdot\!C^{\perp}(\alpha)\big)$}}$}}\\[-5pt]\multicolumn{3}{|@{}l@{}|}{\fwbox{0pt}{\rule[-10pt]{0.0pt}{20.75pt}}\raisebox{-0.15pt}[22pt][21pt]{$\fwbox{355pt}{\hspace{5pt}\displaystyle\scalebox{0.85}{$\displaystyle C(\alpha)\!\equiv\!\left(\!\begin{array}{@{$\,$}c@{$\;\,\,$}c@{$\;\,\,$}c@{$\;\,\,$}c@{$\;\,\,$}c@{$\;\,\,$}c@{$\,\,\,$}}\alpha_6\,\alpha_8&\alpha_1&1&\alpha_6&\alpha_1\,\alpha_7&{\color{dim}0}\\\alpha_8&{\color{dim}0}&{\color{dim}0}&1&\alpha_5&\alpha_4\\\alpha_3&\alpha_2&{\color{dim}0}&{\color{dim}0}&\alpha_2\,\alpha_7&1\end{array}\!\right)$}}$}}\\\hline\multicolumn{3}{@{}c@{}}{~}\\[-15pt]\end{array}\,\begin{array}{@{}c@{}}\\[-44pt]\hspace{-3.95pt}\fig{-42.9pt}{0.75}{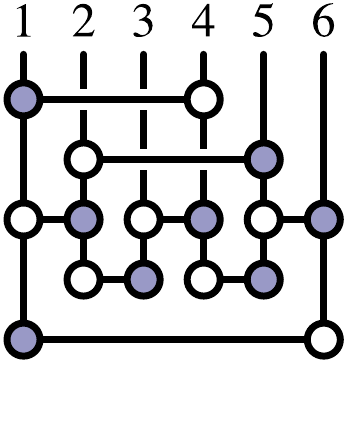}\\[-42pt]\end{array}\nonumber}}

It is worth mentioning that it is not difficult to express $f_8,f_9,f_{10}$ directly as (sums of) rational functions of the kinematical data. To do this, we make use of the global residue theorems from $\Omega_9$, $\Omega_{10}$, and $\Omega_{15}$, respectively. (For the first identity, we also relabel the legs; for the last, we relabel the legs and use parity.) The resulting expressions are: 
\vspace{4pt}\eq{\hspace{-165pt}\fwboxL{400pt}{\begin{array}{r@{$\,$}c@{$\,$}c@{$\,$}c@{$\,$}c@{$\,$}c@{$\,$}c}f_{\fwbox{6.5pt}{8}}(1\,2\,3\,4\,5\,6)\equiv&\phantom{+}&f_{\fwbox{6.5pt}{1}}(3\,2\,1\,6\,5\,4)&-&f_{\fwbox{6.5pt}{1}}(4\,5\,6\,3\,2\,1)&-&f_{\fwbox{6.5pt}{2}}(3\,6\,2\,1\,4\,5)\\[5pt]&-&\widetilde{f}_{\fwbox{6.5pt}{2}}(5\,3\,6\,2\,1\,4)&+&f_{\fwbox{6.5pt}{5}}(4\,3\,5\,6\,2\,1)\end{array}\,.}\hspace{-200pt}}
\eq{\hspace{-165pt}\fwboxL{400pt}{\begin{array}{r@{$\,$}c@{$\,$}c@{$\,$}c@{$\,$}c@{$\,$}c@{$\,$}c}f_{\fwbox{6.5pt}{9}}(1\,2\,3\,4\,5\,6)\equiv&-&f_{\fwbox{6.5pt}{2}}(1\,2\,3\,4\,5\,6)&+&f_{\fwbox{6.5pt}{2}}(2\,3\,6\,1\,4\,5)&-&\widetilde{f}_{\fwbox{6.5pt}{3}}(1\,3\,4\,5\,6\,2)\\[5pt]&+&\widetilde{f}_{\fwbox{6.5pt}{3}}(6\,2\,5\,4\,1\,3)&-&f_{\fwbox{6.5pt}{5}}(4\,2\,5\,6\,3\,1)&+&f_{\fwbox{6.5pt}{5}}(5\,3\,4\,1\,2\,6)\end{array}\,.}\hspace{-200pt}}
\eq{\hspace{-165pt}\fwboxL{400pt}{\begin{array}{r@{$\,$}c@{$\,$}c@{$\,$}c@{$\,$}c@{$\,$}c@{$\,$}c}f_{\fwbox{6.5pt}{\!10}}(1\,2\,3\,4\,5\,6)\equiv&-&\widetilde{f}_{\fwbox{6.5pt}{1}}(6\,4\,1\,2\,5\,3)&-&f_{\fwbox{6.5pt}{2}}(6\,3\,4\,1\,2\,5)&-&\widetilde{f}_{\fwbox{6.5pt}{2}}(5\,6\,3\,4\,1\,2)\\[5pt]&-&\widetilde{f}_{\fwbox{6.5pt}{5}}(2\,6\,1\,4\,5\,3)&-&\widetilde{f}_{\fwbox{6.5pt}{5}}(6\,5\,4\,3\,2\,1)&-&\widetilde{f}_{\fwbox{6.5pt}{7}}(5\,2\,3\,6\,1\,4)\end{array}\,.}\hspace{-200pt}}
Here, we have used $\widetilde{f}_i$ to denote the parity conjugate of the on-shell function $f_i$. It is worth making clear that the above expressions eliminate any ambiguity about the overall signs of these functions---being determined by the signs (conventionally fixed) for the rational functions $f_i$ given above. 

In fact, every top-dimensional form $\Omega_i$ (except $\Omega_{24}$) generates an identity among the on-shell functions $f_i$ via the global residue theorem. And it is worthwhile to give these identities explicitly (with signs fixed by the definitions given above):
\eq{\hspace{-225pt}\fwboxL{400pt}{\begin{array}{r@{$\,$}c@{$\,$}c@{$\,$}c@{$\,$}c@{$\,$}c@{$\,$}c@{$\,$}c@{$\,$}c}\oint\!\Omega_{\fwbox{8pt}{1}}=0=&\phantom{+}&f_{\fwbox{6.5pt}{1}}(1\,2\,3\,4\,5\,6)&-&f_{\fwbox{6.5pt}{1}}(2\,3\,4\,5\,6\,1)&+&f_{\fwbox{6.5pt}{1}}(3\,4\,5\,6\,1\,2)&-&f_{\fwbox{6.5pt}{1}}(4\,5\,6\,1\,2\,3)\\[5pt]&+&f_{\fwbox{6.5pt}{1}}(1\,6\,5\,4\,3\,2)&-&f_{\fwbox{6.5pt}{1}}(2\,1\,6\,5\,4\,3)\end{array}}\hspace{-200pt}\label{grt_for_omega_1}}
\eq{\hspace{-225pt}\fwboxL{400pt}{\begin{array}{r@{$\,$}c@{$\,$}c@{$\,$}c@{$\,$}c@{$\,$}c@{$\,$}c@{$\,$}c@{$\,$}c}\oint\!\Omega_{\fwbox{8pt}{2}}=0=&\phantom{+}&f_{\fwbox{6.5pt}{2}}(1\,2\,3\,4\,5\,6)&+&\widetilde{f}_{\fwbox{6.5pt}{3}}(4\,2\,5\,6\,1\,3)&+&\widetilde{f}_{\fwbox{6.5pt}{3}}(4\,2\,3\,1\,6\,5)&+&f_{\fwbox{6.5pt}{1}}(2\,4\,3\,1\,6\,5)\\[5pt]&+&f_{\fwbox{6.5pt}{1}}(2\,4\,5\,6\,1\,3)&+&f_{\fwbox{6.5pt}{2}}(5\,2\,6\,1\,3\,4)&+&\widetilde{f}_{\fwbox{6.5pt}{2}}(1\,2\,6\,5\,4\,3)\end{array}}\hspace{-200pt}\label{grt_for_omega_2}}
\eq{\hspace{-225pt}\fwboxL{400pt}{\begin{array}{r@{$\,$}c@{$\,$}c@{$\,$}c@{$\,$}c@{$\,$}c@{$\,$}c@{$\,$}c@{$\,$}c}\oint\!\Omega_{\fwbox{8pt}{3}}=0=&\phantom{+}&f_{\fwbox{6.5pt}{2}}(1\,2\,3\,4\,5\,6)&+&f_{\fwbox{6.5pt}{4}}(3\,1\,5\,4\,2\,6)&-&f_{\fwbox{6.5pt}{7}}(5\,1\,6\,2\,3\,4)&-&f_{\fwbox{6.5pt}{1}}(2\,4\,3\,1\,5\,6)\\[5pt]&+&f_{\fwbox{6.5pt}{2}}(2\,3\,6\,1\,5\,4)&+&\widetilde{f}_{\fwbox{6.5pt}{2}}(1\,6\,3\,2\,4\,5)&-&f_{\fwbox{6.5pt}{4}}(2\,6\,5\,1\,3\,4)&+&\widetilde{f}_{\fwbox{6.5pt}{2}}(2\,1\,6\,5\,4\,3)\end{array}}\hspace{-200pt}\label{grt_for_omega_3}}
\eq{\hspace{-225pt}\fwboxL{400pt}{\begin{array}{r@{$\,$}c@{$\,$}c@{$\,$}c@{$\,$}c@{$\,$}c@{$\,$}c@{$\,$}c@{$\,$}c}\oint\!\Omega_{\fwbox{8pt}{4}}=0=&\phantom{+}&f_{\fwbox{6.5pt}{3}}(1\,2\,3\,4\,5\,6)&-&\widetilde{f}_{\fwbox{6.5pt}{2}}(3\,6\,4\,5\,2\,1)&+&f_{\fwbox{6.5pt}{4}}(1\,3\,4\,2\,5\,6)&-&f_{\fwbox{6.5pt}{3}}(1\,3\,6\,4\,5\,2)\\[5pt]&+&\widetilde{f}_{\fwbox{6.5pt}{2}}(3\,2\,4\,5\,6\,1)&+&f_{\fwbox{6.5pt}{3}}(5\,2\,4\,3\,1\,6)&-&f_{\fwbox{6.5pt}{4}}(3\,4\,5\,2\,1\,6)&-&f_{\fwbox{6.5pt}{3}}(5\,4\,6\,3\,1\,2)\end{array}}\hspace{-200pt}\label{grt_for_omega_4}}
\eq{\hspace{-225pt}\fwboxL{400pt}{\begin{array}{r@{$\,$}c@{$\,$}c@{$\,$}c@{$\,$}c@{$\,$}c@{$\,$}c@{$\,$}c@{$\,$}c}\oint\!\Omega_{\fwbox{8pt}{5}}=0=&\phantom{+}&f_{\fwbox{6.5pt}{4}}(1\,2\,3\,4\,5\,6)&-&f_{\fwbox{6.5pt}{4}}(1\,2\,5\,4\,3\,6)&-&f_{\fwbox{6.5pt}{4}}(1\,4\,3\,2\,5\,6)&+&f_{\fwbox{6.5pt}{4}}(1\,6\,3\,2\,5\,4)\\[5pt]&+&f_{\fwbox{6.5pt}{4}}(1\,4\,5\,2\,3\,6)&-&f_{\fwbox{6.5pt}{4}}(1\,6\,5\,2\,3\,4)&+&f_{\fwbox{6.5pt}{4}}(3\,2\,5\,4\,1\,6)&-&f_{\fwbox{6.5pt}{4}}(3\,4\,5\,2\,1\,6)\\[5pt]&+&f_{\fwbox{6.5pt}{4}}(3\,6\,5\,2\,1\,4)\end{array}}\hspace{-200pt}\label{grt_for_omega_5}}
\eq{\hspace{-225pt}\fwboxL{400pt}{\begin{array}{r@{$\,$}c@{$\,$}c@{$\,$}c@{$\,$}c@{$\,$}c@{$\,$}c@{$\,$}c@{$\,$}c}\oint\!\Omega_{\fwbox{8pt}{6}}=0=&\phantom{+}&f_{\fwbox{6.5pt}{3}}(1\,2\,3\,4\,5\,6)&+&\widetilde{f}_{\fwbox{6.5pt}{2}}(3\,6\,4\,5\,1\,2)&+&\widetilde{f}_{\fwbox{6.5pt}{2}}(2\,6\,4\,5\,1\,3)&-&f_{\fwbox{6.5pt}{7}}(5\,1\,6\,2\,3\,4)\\[5pt]&+&f_{\fwbox{6.5pt}{3}}(4\,2\,3\,1\,5\,6)&+&\widetilde{f}_{\fwbox{6.5pt}{6}}(1\,5\,4\,2\,3\,6)&+&\widetilde{f}_{\fwbox{6.5pt}{2}}(1\,6\,3\,2\,4\,5)&+&\widetilde{f}_{\fwbox{6.5pt}{2}}(1\,6\,2\,3\,4\,5)\\[5pt]&-&f_{\fwbox{6.5pt}{7}}(5\,4\,6\,2\,3\,1)\end{array}}\hspace{-200pt}\label{grt_for_omega_6}}
\eq{\hspace{-225pt}\fwboxL{400pt}{\begin{array}{r@{$\,$}c@{$\,$}c@{$\,$}c@{$\,$}c@{$\,$}c@{$\,$}c@{$\,$}c@{$\,$}c}\oint\!\Omega_{\fwbox{8pt}{7}}=0=&\phantom{+}&f_{\fwbox{6.5pt}{3}}(1\,2\,3\,4\,5\,6)&-&f_{\fwbox{6.5pt}{1}}(2\,1\,6\,5\,4\,3)&-&f_{\fwbox{6.5pt}{1}}(3\,1\,6\,5\,4\,2)&+&\widetilde{f}_{\fwbox{6.5pt}{3}}(4\,2\,3\,1\,6\,5)\\[5pt]&-&f_{\fwbox{6.5pt}{3}}(4\,2\,3\,1\,6\,5)&+&f_{\fwbox{6.5pt}{1}}(2\,4\,5\,6\,1\,3)&+&f_{\fwbox{6.5pt}{1}}(3\,4\,5\,6\,1\,2)&-&\widetilde{f}_{\fwbox{6.5pt}{3}}(1\,2\,3\,4\,5\,6)\end{array}}\hspace{-200pt}\label{grt_for_omega_7}}
\eq{\hspace{-225pt}\fwboxL{400pt}{\begin{array}{r@{$\,$}c@{$\,$}c@{$\,$}c@{$\,$}c@{$\,$}c@{$\,$}c@{$\,$}c@{$\,$}c}\oint\!\Omega_{\fwbox{8pt}{8}}=0=&\phantom{+}&f_{\fwbox{6.5pt}{7}}(1\,2\,3\,4\,5\,6)&+&\widetilde{f}_{\fwbox{6.5pt}{3}}(3\,4\,5\,6\,1\,2)&+&\widetilde{f}_{\fwbox{6.5pt}{3}}(2\,4\,5\,6\,1\,3)&-&\widetilde{f}_{\fwbox{6.5pt}{3}}(5\,2\,3\,1\,6\,4)\\[5pt]&-&\widetilde{f}_{\fwbox{6.5pt}{3}}(4\,2\,3\,1\,6\,5)&+&f_{\fwbox{6.5pt}{3}}(4\,2\,3\,1\,6\,5)&+&f_{\fwbox{6.5pt}{3}}(5\,2\,3\,1\,6\,4)&-&f_{\fwbox{6.5pt}{3}}(2\,4\,5\,6\,1\,3)\\[5pt]&-&f_{\fwbox{6.5pt}{3}}(3\,4\,5\,6\,1\,2)&-&f_{\fwbox{6.5pt}{7}}(6\,4\,5\,2\,3\,1)\end{array}}\hspace{-200pt}\label{grt_for_omega_8}}
\eq{\hspace{-225pt}\fwboxL{400pt}{\begin{array}{r@{$\,$}c@{$\,$}c@{$\,$}c@{$\,$}c@{$\,$}c@{$\,$}c@{$\,$}c@{$\,$}c}\oint\!\Omega_{\fwbox{8pt}{9}}=0=&\phantom{+}&f_{\fwbox{6.5pt}{5}}(1\,2\,3\,4\,5\,6)&-&f_{\fwbox{6.5pt}{1}}(1\,3\,4\,2\,5\,6)&-&\widetilde{f}_{\fwbox{6.5pt}{2}}(3\,2\,4\,5\,6\,1)&-&f_{\fwbox{6.5pt}{2}}(2\,4\,5\,6\,1\,3)\\[5pt]&+&f_{\fwbox{6.5pt}{1}}(2\,5\,6\,4\,3\,1)&-&f_{\fwbox{6.5pt}{8}}(6\,5\,2\,1\,3\,4)\end{array}}\hspace{-200pt}\label{grt_for_omega_9}}
\eq{\hspace{-225pt}\fwboxL{400pt}{\begin{array}{r@{$\,$}c@{$\,$}c@{$\,$}c@{$\,$}c@{$\,$}c@{$\,$}c@{$\,$}c@{$\,$}c}\oint\!\Omega_{\fwbox{8pt}{10}}=0=&\phantom{+}&f_{\fwbox{6.5pt}{2}}(1\,2\,3\,4\,5\,6)&-&\widetilde{f}_{\fwbox{6.5pt}{3}}(6\,2\,5\,4\,1\,3)&-&f_{\fwbox{6.5pt}{2}}(2\,3\,6\,1\,4\,5)&+&f_{\fwbox{6.5pt}{5}}(4\,2\,5\,6\,3\,1)\\[5pt]&+&\widetilde{f}_{\fwbox{6.5pt}{3}}(1\,3\,4\,5\,6\,2)&-&f_{\fwbox{6.5pt}{5}}(5\,3\,4\,1\,2\,6)&-&f_{\fwbox{6.5pt}{9}}(1\,2\,3\,4\,5\,6)\end{array}}\hspace{-200pt}\label{grt_for_omega_10}}
\eq{\hspace{-225pt}\fwboxL{400pt}{\begin{array}{r@{$\,$}c@{$\,$}c@{$\,$}c@{$\,$}c@{$\,$}c@{$\,$}c@{$\,$}c@{$\,$}c}\oint\!\Omega_{\fwbox{8pt}{11}}=0=&\phantom{+}&f_{\fwbox{6.5pt}{5}}(1\,2\,3\,4\,5\,6)&-&f_{\fwbox{6.5pt}{1}}(3\,1\,6\,5\,2\,4)&-&\widetilde{f}_{\fwbox{6.5pt}{3}}(4\,2\,3\,1\,6\,5)&-&f_{\fwbox{6.5pt}{1}}(2\,4\,3\,1\,6\,5)\\[5pt]&-&f_{\fwbox{6.5pt}{3}}(2\,4\,5\,6\,1\,3)&-&f_{\fwbox{6.5pt}{4}}(2\,5\,6\,1\,3\,4)&+&f_{\fwbox{6.5pt}{8}}(6\,4\,3\,1\,2\,5)\end{array}}\hspace{-200pt}\label{grt_for_omega_11}}

\eq{\hspace{-225pt}\fwboxL{400pt}{\begin{array}{r@{$\,$}c@{$\,$}c@{$\,$}c@{$\,$}c@{$\,$}c@{$\,$}c@{$\,$}c@{$\,$}c}\oint\!\Omega_{\fwbox{8pt}{12}}=0=&\phantom{+}&f_{\fwbox{6.5pt}{6}}(1\,2\,3\,4\,5\,6)&-&\widetilde{f}_{\fwbox{6.5pt}{3}}(3\,5\,6\,2\,1\,4)&+&f_{\fwbox{6.5pt}{2}}(1\,5\,4\,6\,3\,2)&+&f_{\fwbox{6.5pt}{2}}(1\,6\,4\,5\,3\,2)\\[5pt]&-&f_{\fwbox{6.5pt}{5}}(3\,4\,5\,6\,1\,2)&-&f_{\fwbox{6.5pt}{5}}(3\,4\,6\,5\,1\,2)&+&f_{\fwbox{6.5pt}{3}}(3\,5\,6\,4\,1\,2)&+&f_{\fwbox{6.5pt}{9}}(1\,5\,6\,4\,3\,2)\end{array}}\hspace{-200pt}\label{grt_for_omega_12}}
\eq{\hspace{-225pt}\fwboxL{400pt}{\begin{array}{r@{$\,$}c@{$\,$}c@{$\,$}c@{$\,$}c@{$\,$}c@{$\,$}c@{$\,$}c@{$\,$}c}\oint\!\Omega_{\fwbox{8pt}{13}}=0=&\phantom{+}&f_{\fwbox{6.5pt}{7}}(1\,2\,3\,4\,5\,6)&+&f_{\fwbox{6.5pt}{4}}(2\,1\,4\,3\,5\,6)&+&f_{\fwbox{6.5pt}{4}}(3\,1\,4\,2\,5\,6)&+&f_{\fwbox{6.5pt}{7}}(4\,1\,6\,2\,3\,5)\\[5pt]&-&f_{\fwbox{6.5pt}{3}}(5\,2\,3\,1\,4\,6)&+&f_{\fwbox{6.5pt}{5}}(2\,4\,5\,6\,1\,3)&+&f_{\fwbox{6.5pt}{5}}(3\,4\,5\,6\,1\,2)&-&\widetilde{f}_{\fwbox{6.5pt}{3}}(1\,2\,3\,5\,6\,4)\\[5pt]&+&f_{\fwbox{6.5pt}{9}}(1\,2\,3\,4\,5\,6)\end{array}}\hspace{-200pt}\label{grt_for_omega_13}}
\eq{\hspace{-225pt}\fwboxL{400pt}{\begin{array}{r@{$\,$}c@{$\,$}c@{$\,$}c@{$\,$}c@{$\,$}c@{$\,$}c@{$\,$}c@{$\,$}c}\oint\!\Omega_{\fwbox{8pt}{14}}=0=&\phantom{+}&f_{\fwbox{6.5pt}{5}}(1\,2\,3\,4\,5\,6)&-&f_{\fwbox{6.5pt}{2}}(2\,1\,4\,3\,5\,6)&-&\widetilde{f}_{\fwbox{6.5pt}{2}}(2\,6\,4\,1\,3\,5)&-&\widetilde{f}_{\fwbox{6.5pt}{2}}(2\,3\,4\,5\,6\,1)\\[5pt]&-&f_{\fwbox{6.5pt}{2}}(2\,5\,4\,6\,1\,3)&+&f_{\fwbox{6.5pt}{5}}(5\,2\,6\,4\,1\,3)&+&\widetilde{f}_{\fwbox{6.5pt}{\!10}}(1\,3\,4\,6\,5\,2)\end{array}}\hspace{-200pt}\label{grt_for_omega_14}}
\eq{\hspace{-225pt}\fwboxL{400pt}{\begin{array}{r@{$\,$}c@{$\,$}c@{$\,$}c@{$\,$}c@{$\,$}c@{$\,$}c@{$\,$}c@{$\,$}c}\oint\!\Omega_{\fwbox{8pt}{15}}=0=&\phantom{+}&f_{\fwbox{6.5pt}{5}}(1\,2\,3\,4\,5\,6)&+&\widetilde{f}_{\fwbox{6.5pt}{2}}(2\,6\,4\,3\,1\,5)&+&f_{\fwbox{6.5pt}{7}}(5\,1\,6\,2\,3\,4)&+&f_{\fwbox{6.5pt}{1}}(2\,4\,3\,1\,5\,6)\\[5pt]&+&f_{\fwbox{6.5pt}{5}}(2\,5\,4\,6\,1\,3)&+&f_{\fwbox{6.5pt}{2}}(5\,2\,6\,4\,3\,1)&-&\widetilde{f}_{\fwbox{6.5pt}{\!10}}(3\,1\,6\,4\,5\,2)\end{array}}\hspace{-200pt}\label{grt_for_omega_15}}

\eq{\hspace{-225pt}\fwboxL{400pt}{\begin{array}{r@{$\,$}c@{$\,$}c@{$\,$}c@{$\,$}c@{$\,$}c@{$\,$}c@{$\,$}c@{$\,$}c}\oint\!\Omega_{\fwbox{8pt}{16}}=0=&\phantom{+}&f_{\fwbox{6.5pt}{5}}(1\,2\,4\,3\,6\,5)&-&\widetilde{f}_{\fwbox{6.5pt}{2}}(2\,4\,3\,6\,5\,1)&-&f_{\fwbox{6.5pt}{2}}(2\,6\,3\,5\,1\,4)&+&f_{\fwbox{6.5pt}{8}}(1\,2\,3\,4\,5\,6)\\[5pt]&-&f_{\fwbox{6.5pt}{\!10}}(2\,1\,4\,3\,5\,6)\end{array}}\hspace{-200pt}\label{grt_for_omega_16}}
\eq{\hspace{-225pt}\fwboxL{400pt}{\begin{array}{r@{$\,$}c@{$\,$}c@{$\,$}c@{$\,$}c@{$\,$}c@{$\,$}c@{$\,$}c@{$\,$}c}\oint\!\Omega_{\fwbox{8pt}{17}}=0=&\phantom{+}&f_{\fwbox{6.5pt}{5}}(1\,2\,3\,4\,5\,6)&+&f_{\fwbox{6.5pt}{1}}(3\,1\,4\,2\,5\,6)&+&f_{\fwbox{6.5pt}{5}}(2\,4\,5\,6\,1\,3)&+&\widetilde{f}_{\fwbox{6.5pt}{3}}(1\,3\,4\,6\,5\,2)\\[5pt]&-&f_{\fwbox{6.5pt}{9}}(6\,2\,3\,5\,4\,1)&+&f_{\fwbox{6.5pt}{\!10}}(1\,2\,3\,4\,5\,6)\end{array}}\hspace{-200pt}\label{grt_for_omega_17}}
\eq{\hspace{-225pt}\fwboxL{400pt}{\begin{array}{r@{$\,$}c@{$\,$}c@{$\,$}c@{$\,$}c@{$\,$}c@{$\,$}c@{$\,$}c@{$\,$}c}\oint\!\Omega_{\fwbox{8pt}{18}}=0=&\phantom{+}&f_{\fwbox{6.5pt}{5}}(1\,2\,4\,3\,6\,5)&+&f_{\fwbox{6.5pt}{5}}(1\,3\,4\,2\,6\,5)&-&f_{\fwbox{6.5pt}{6}}(1\,5\,6\,2\,3\,4)&-&\widetilde{f}_{\fwbox{6.5pt}{6}}(4\,1\,5\,2\,3\,6)\\[5pt]&+&f_{\fwbox{6.5pt}{9}}(1\,2\,3\,4\,5\,6)&-&f_{\fwbox{6.5pt}{\!10}}(3\,1\,4\,2\,5\,6)&-&f_{\fwbox{6.5pt}{\!10}}(2\,1\,4\,3\,5\,6)\end{array}}\hspace{-200pt}\label{grt_for_omega_18}}
\eq{\hspace{-225pt}\fwboxL{400pt}{\begin{array}{r@{$\,$}c@{$\,$}c@{$\,$}c@{$\,$}c@{$\,$}c@{$\,$}c@{$\,$}c@{$\,$}c}\oint\!\Omega_{\fwbox{8pt}{19}}=0=&\phantom{+}&f_{\fwbox{6.5pt}{5}}(3\,1\,2\,5\,6\,4)&+&f_{\fwbox{6.5pt}{5}}(4\,2\,1\,5\,6\,3)&+&f_{\fwbox{6.5pt}{1}}(1\,5\,2\,3\,6\,4)&+&f_{\fwbox{6.5pt}{2}}(4\,1\,6\,3\,2\,5)\\[5pt]&+&f_{\fwbox{6.5pt}{2}}(3\,2\,6\,4\,1\,5)&+&f_{\fwbox{6.5pt}{7}}(6\,3\,4\,1\,2\,5)&+&f_{\fwbox{6.5pt}{\!10}}(2\,1\,4\,3\,5\,6)&+&f_{\fwbox{6.5pt}{\!10}}(1\,2\,3\,4\,5\,6)\end{array}}\hspace{-200pt}\label{grt_for_omega_19}}
\eq{\hspace{-225pt}\fwboxL{400pt}{\begin{array}{r@{$\,$}c@{$\,$}c@{$\,$}c@{$\,$}c@{$\,$}c@{$\,$}c@{$\,$}c@{$\,$}c}\oint\!\Omega_{\fwbox{8pt}{20}}=0=&\phantom{+}&f_{\fwbox{6.5pt}{\!10}}(1\,2\,3\,4\,5\,6)&+&f_{\fwbox{6.5pt}{\!10}}(3\,2\,1\,6\,5\,4)&+&f_{\fwbox{6.5pt}{\!10}}(1\,6\,5\,4\,3\,2)\end{array}}\hspace{-200pt}\label{grt_for_omega_20}}
\eq{\hspace{-225pt}\fwboxL{400pt}{\begin{array}{r@{$\,$}c@{$\,$}c@{$\,$}c@{$\,$}c@{$\,$}c@{$\,$}c@{$\,$}c@{$\,$}c}\oint\!\Omega_{\fwbox{8pt}{21}}=0=&\phantom{+}&f_{\fwbox{6.5pt}{5}}(1\,2\,3\,4\,5\,6)&-&f_{\fwbox{6.5pt}{5}}(4\,6\,1\,3\,2\,5)&-&f_{\fwbox{6.5pt}{2}}(2\,6\,5\,3\,1\,4)&+&\widetilde{f}_{\fwbox{6.5pt}{2}}(1\,4\,3\,6\,5\,2)\\[5pt]&-&f_{\fwbox{6.5pt}{\!10}}(3\,5\,4\,1\,6\,2)&+&f_{\fwbox{6.5pt}{\!10}}(1\,2\,3\,4\,5\,6)\end{array}}\hspace{-200pt}\label{grt_for_omega_21}}
\eq{\hspace{-225pt}\fwboxL{400pt}{\begin{array}{r@{$\,$}c@{$\,$}c@{$\,$}c@{$\,$}c@{$\,$}c@{$\,$}c@{$\,$}c@{$\,$}c}\oint\!\Omega_{\fwbox{8pt}{22}}=0=&\phantom{+}&f_{\fwbox{6.5pt}{5}}(1\,3\,2\,5\,4\,6)&+&\widetilde{f}_{\fwbox{6.5pt}{3}}(3\,5\,6\,2\,1\,4)&+&f_{\fwbox{6.5pt}{1}}(1\,2\,5\,3\,6\,4)&+&f_{\fwbox{6.5pt}{5}}(4\,1\,6\,3\,2\,5)\\[5pt]&+&f_{\fwbox{6.5pt}{5}}(4\,6\,1\,2\,3\,5)&+&f_{\fwbox{6.5pt}{5}}(3\,4\,5\,6\,1\,2)&+&f_{\fwbox{6.5pt}{3}}(6\,3\,4\,1\,2\,5)&+&f_{\fwbox{6.5pt}{1}}(5\,3\,6\,4\,1\,2)\\[5pt]&-&f_{\fwbox{6.5pt}{\!10}}(1\,2\,3\,4\,5\,6)\end{array}}\hspace{-200pt}\label{grt_for_omega_22}}
\eq{\hspace{-225pt}\fwboxL{400pt}{\begin{array}{r@{$\,$}c@{$\,$}c@{$\,$}c@{$\,$}c@{$\,$}c@{$\,$}c@{$\,$}c@{$\,$}c}\oint\!\Omega_{\fwbox{8pt}{23}}=0=&\phantom{+}&f_{\fwbox{6.5pt}{7}}(1\,2\,5\,3\,4\,6)&-&f_{\fwbox{6.5pt}{7}}(2\,1\,6\,3\,4\,5)&-&f_{\fwbox{6.5pt}{7}}(1\,3\,4\,2\,5\,6)&+&f_{\fwbox{6.5pt}{7}}(3\,1\,6\,2\,5\,4)\\[5pt]&+&f_{\fwbox{6.5pt}{7}}(4\,1\,6\,2\,5\,3)&-&f_{\fwbox{6.5pt}{7}}(5\,1\,6\,3\,4\,2)&+&f_{\fwbox{6.5pt}{7}}(2\,3\,4\,1\,6\,5)&-&f_{\fwbox{6.5pt}{7}}(3\,2\,5\,1\,6\,4)\\[5pt]&-&f_{\fwbox{6.5pt}{7}}(4\,2\,5\,1\,6\,3)&+&f_{\fwbox{6.5pt}{7}}(6\,2\,5\,3\,4\,1)&+&f_{\fwbox{6.5pt}{7}}(5\,3\,4\,1\,6\,2)&-&f_{\fwbox{6.5pt}{7}}(6\,3\,4\,2\,5\,1)\end{array}}\hspace{-200pt}\label{grt_for_omega_23}}
%
\eq{\hspace{-225pt}\fwboxL{400pt}{\begin{array}{r@{$\,$}c@{$\,$}c@{$\,$}c@{$\,$}c@{$\,$}c@{$\,$}c@{$\,$}c@{$\,$}c}\oint\!\Omega_{\fwbox{8pt}{24}}=0\equiv&\phantom{+}&f_{\fwbox{6.5pt}{5}}(4\,5\,1\,2\,6\,3)&-&f_{\fwbox{6.5pt}{5}}(4\,5\,1\,2\,6\,3)&+&f_{\fwbox{6.5pt}{5}}(4\,6\,1\,3\,5\,2)&-&f_{\fwbox{6.5pt}{5}}(4\,6\,1\,3\,5\,2)\\[5pt]&+&f_{\fwbox{6.5pt}{5}}(5\,4\,2\,1\,6\,3)&-&f_{\fwbox{6.5pt}{5}}(5\,4\,2\,1\,6\,3)&+&f_{\fwbox{6.5pt}{5}}(5\,6\,2\,3\,4\,1)&-&f_{\fwbox{6.5pt}{5}}(5\,6\,2\,3\,4\,1)\\[5pt]&+&f_{\fwbox{6.5pt}{5}}(6\,5\,3\,2\,4\,1)&-&f_{\fwbox{6.5pt}{5}}(6\,5\,3\,2\,4\,1)&+&f_{\fwbox{6.5pt}{5}}(6\,4\,3\,1\,5\,2)&-&f_{\fwbox{6.5pt}{5}}(6\,4\,3\,1\,5\,2)\end{array}}\hspace{-200pt}\label{grt_for_omega_24}}

\vspace{4pt}
Using these identities, one can eliminate all appearances of $f_8,f_9$, and $f_{10}$. From the seven rational functions, we may expect $10,\!080$ different functions through permutations and parity; however, only $3,\!000$ of these are distinct. The identities above span a space of $2,\!566$ independent identities, leaving a space of only $434$ linearly-independent combinations of on-shell functions. 

Notice that the last identity for $\Omega_{24}$, (\ref{grt_for_omega_24}), is trivial: the 12 on-shell functions along its boundary come in 6, mutually canceling pairs. This follows from the interpretation that each removable edge should contribute one configuration to the boundary---despite the appearance of move-equivalent configurations.

\newpage
\vspace{-12pt}\section{Representative Lower Dimension On-Shell Varieties of $G(3,6)$}\label{appendix:higher_singularities}\vspace{-6pt}
For the sake of reference and completeness, in this appendix we provide representative on-shell diagrams for each class of inequivalent on-shell varieties of dimension 7 and 6. On-shell varieties of lower dimension are all planar, representatives of which can easily be generated by the {\sc Mathematica} package {\tt positroids} described in \cite{Bourjaily:2012gy}.

At dimension 7, there are seven inequivalent on-shell varieties---three of which are planar varieties. Representatives of these are as follows:
\eq{\begin{array}{c}\fig{-42.9pt}{0.75}{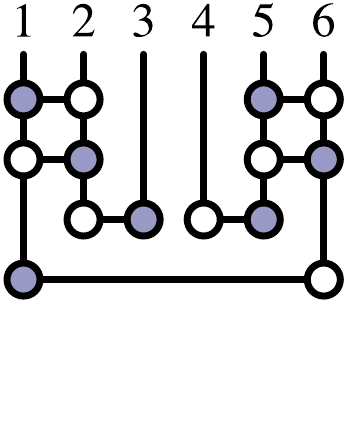}\;\;\fig{-42.9pt}{0.75}{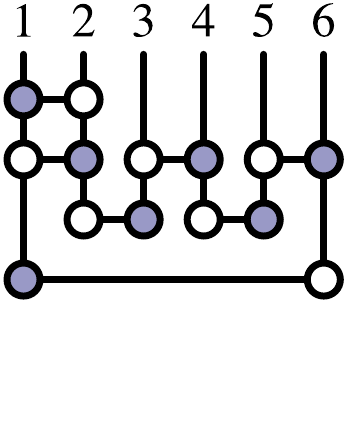}\;\;\fig{-42.9pt}{0.75}{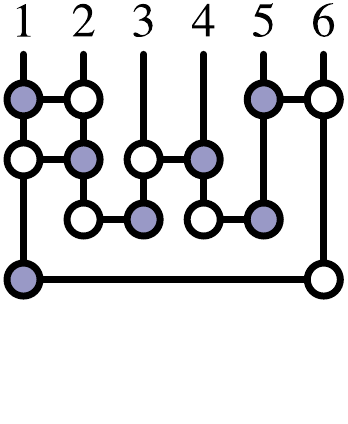}\;\;\fig{-42.9pt}{0.75}{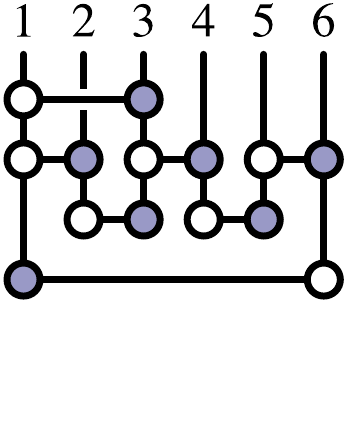}\\[-20pt]
\fig{-42.9pt}{0.75}{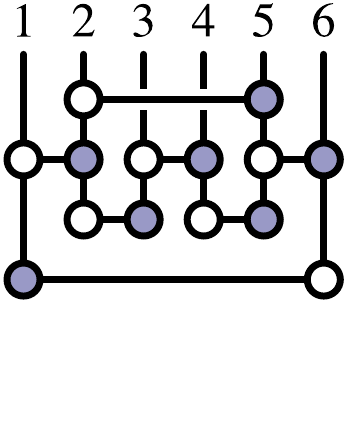}\;\;\fig{-42.9pt}{0.75}{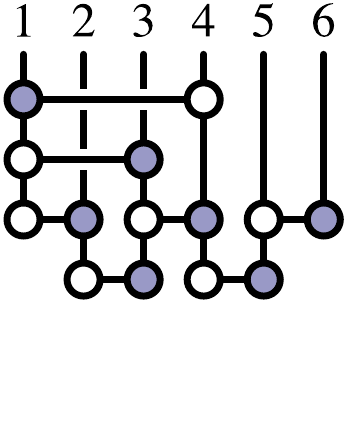}\;\;\fig{-42.9pt}{0.75}{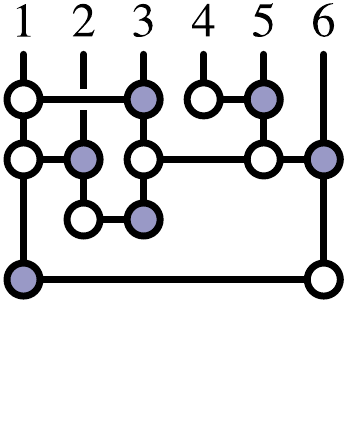}\\[-24pt]\end{array}\label{7_dim_varieties}}

At dimension 6, there are six inequivalent on-shell varieties---all but one of which are planar. The non-planar variety corresponds to the unique non-planar variety in $G(2,5)$, with an additional zero-dimensional component corresponding to a `hanging' leg in the on-shell diagram. Representatives of these six on-shell varieties are:
\eq{\begin{array}{c}\fig{-42.9pt}{0.75}{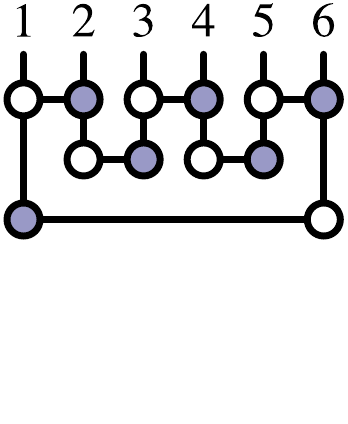}\;\;\fig{-42.9pt}{0.75}{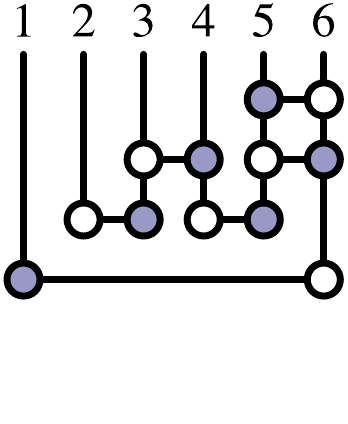}\;\;\fig{-42.9pt}{0.75}{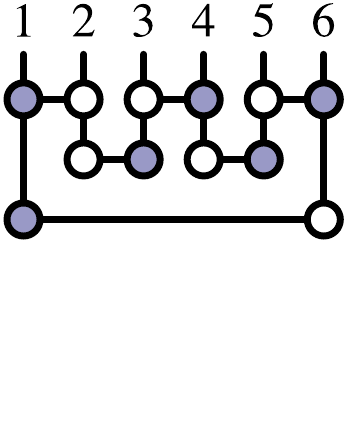}\\[-20pt]
\fig{-42.9pt}{0.75}{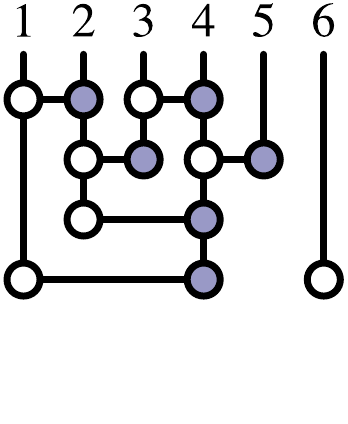}\;\;\fig{-42.9pt}{0.75}{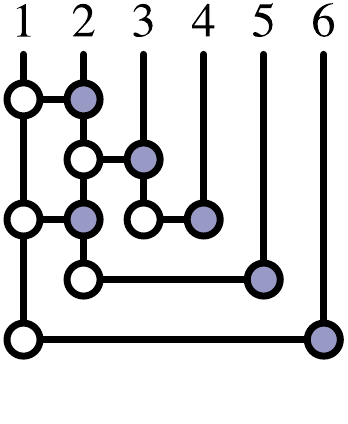}\;\;\fig{-42.9pt}{0.75}{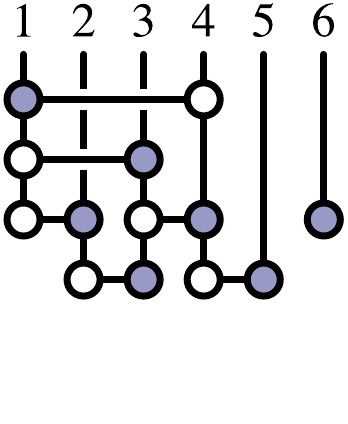}\\[-16pt]\end{array}\label{6_dim_varieties}}

All on-shell varieties below dimension 6 admit a planar embedding on the disk, and therefore correspond to familiar positroids. The numbers of such varieties are listed in \mbox{Table \ref{table_of_strata}}. We should remind the reader that we count only those classes of varieties inequivalent under parity and relabeling of the external legs.

\newpage
\providecommand{\href}[2]{#2}\begingroup\raggedright\endgroup

\end{document}